\documentclass[12pt]{article}
\usepackage{amsmath,amsfonts,amssymb,mathrsfs,cite}
\usepackage{color,graphicx,shortvrb,epsfig}
\usepackage{latexsym}
\usepackage[a4paper]{geometry}
\usepackage[in]{fullpage}
\newlength{\myem}
\newcounter{mysubequation}[equation]

\newcommand{\capdef}{}
\newcommand{\mycaption}[2][\capdef]{\renewcommand{\capdef}{#2}%
        \caption[#1]{{\footnotesize #2}}}
\makeatletter
\renewcommand{\section}{\@startsection{section}{1}{0em}{-\baselineskip}%
{\baselineskip}{\normalfont\large\bfseries}}
\renewcommand{\subsection}%
{\@startsection{subsection}{2}{0em}{-0.7\baselineskip}%
{0.7\baselineskip}{\normalfont\bfseries}}
\makeatother

\newcommand{\ba}{\begin{array}}
\newcommand{\ea}{\end{array}}
\newcommand{\bd}{\begin{displaymath}}
\newcommand{\ed}{\end{displaymath}}
\newcommand{\bi}{\begin{itemize}}
\newcommand{\ei}{\end{itemize}}
\newcommand{\benu}{\begin{enumerate}}
\newcommand{\eenu}{\end{enumerate}}
\newcommand{\be}{\begin{equation}}
\newcommand{\ee}{\end{equation}}
\newcommand{\bea}{\begin{eqnarray}}
\newcommand{\eea}{\end{eqnarray}}
\def\lsim{\ \raisebox{-.45ex}{\rlap{$\sim$}} \raisebox{.45ex}{$<$}\ }
\def\gsim{\ \raisebox{-.45ex}{\rlap{$\sim$}} \raisebox{.45ex}{$>$}\ }
\newcommand{\HK}{\mbox{\sf HK~}}
\newcommand{\SK}{\mbox{\sf SK~}}

\newcommand{\INO}{\mbox{\sf INO~}}

\newcommand{\KTOK}{\mbox{\sf K2K~}}

\newcommand{\DCHOOZ}{\mbox{\sf D-CHOOZ~}}

\newcommand{\KamLAND}{\mbox{\sf KamLAND~}}

\newcommand{\THKd}{\mbox{\sf T2K}}
\newcommand{\NOVAd}{\mbox{\sf NO$\nu$A}}
\newcommand{\HKd}{\mbox{\sf HK}}
\newcommand{\SKd}{\mbox{\sf SK}}
\newcommand{\UNOd}{\mbox{\sf UNO}}
\newcommand{\INOd}{\mbox{\sf INO}}
\newcommand{\MEMPHYSd}{\mbox{\sf MEMPHYS}}

\newcommand{\MINOSd}{\mbox{\sf MINOS}}

\newcommand{\CHOOZd}{\mbox{\sf CHOOZ}}

\newcommand{\ie}{{\it i.e.}}

\newcommand{\eg}{{\it e.g.}}

\newcommand{\etc}{{\it etc.}}
\newcommand{\eq}{Eq.}

\newcommand{\fig}{Figure}
\newcommand{\Fig}{Figure}
\newcommand{\figs}{Figures}
\newcommand{\Figs}{Figures}

\newcommand{\Sec}{Section}

\newcommand{\tab}{Table}

\newcommand{\Tab}{Table}

\def\nn{\nonumber}
\def\evsq{{{\rm eV$^2$~}}}

\def\dam{\mathrm{(\Delta m_{31}^2)^m}}

\def\dji{\mathrm{\Delta m_{ji}^2}}
\def\da{\mathrm{\Delta m_{31}^2}}

\def\ds{\mathrm{\Delta m_{21}^2}}
\def\dcp{{\delta_{\mathrm{CP}}}}
\def\stsmall{{\sin^2 2 \theta_{13}}}
\def\stsmallm{\mathrm{\sin^2 2 \theta_{13}^m}}
\def\pee{{{\rm P_{e e}}}}
\def\peem{{{\rm P^m_{e e}}}}
\def\pmutau{{{\rm P_{\mu \tau}}}}
\def\pmue{{{\rm P_{\mu e}}}}
\def\pmuem{{{\rm P^m_{\mu e}}}}
\def\pemum{{{\rm P^m_{e \mu}}}}
\def\pemu{{{\rm P_{e \mu}}}}
\def\pmumu{{{\rm P_{\mu \mu}}}}
\def\pmumum{{{\rm P^m_{\mu \mu}}}}
\def\mtyr{$\mathrm{Mt~yr}~$}
\def\mtyrd{$\mathrm{Mt~yr}$}
\def\signda{{{{\sf{sign}}}}{(\Delta {\rm{m_{31}^2}})}}
\newcommand{\chr}{\mbox{$\breve{\rm C}$erenkov~}}
\begin{document}
%
%
%
%
\begin{flushright}
\end{flushright}
\vspace*{1cm}
\setcounter{footnote}{-1}
    {\begin{center}
    {\LARGE{\bf{Mass Hierarchy determination via future Atmospheric Neutrino Detectors}}}
    \end{center}}
\renewcommand{\thefootnote}{\fnsymbol{footnote}}
\vspace*{1cm}
                {\begin{center}
                {{\bf
                Raj Gandhi $^{a, \,\!\!\!}$
                \footnote[1]{\makebox[0.cm]{}
                \sf raj@mri.ernet.in},
                Pomita Ghoshal $^{a, \,\!\!\!}$
                \footnote[2]{\makebox[0.cm]{}
                \sf pomita@mri.ernet.in},
                Srubabati Goswami $^{a, \,\!\!\!}$
                \footnote[3]{\makebox[0.cm]{}
                \sf sruba@mri.ernet.in},
                Poonam Mehta $^{b, \,\!\!\!}$
                \footnote[4]{\makebox[0.cm]{}
                \sf poonam.mehta@weizmann.ac.il},
                S Uma Sankar $^{c, \,\!\!}$
                \footnote[5]{\makebox[0.cm]{}
                \sf uma@phy.iitb.ac.in},
                Shashank Shalgar $^{a, \,\!\!\!}$
                \footnote[6]{\makebox[0.cm]{}
                \sf shashank@mri.ernet.in}
                }}
                \end{center}}
\vskip 1.2cm
{\small
                \begin{center}
                $^a$ Harish-Chandra Research Institute, Chhatnag Road, \\
                     Jhunsi, Allahabad 211 019, India\\[4mm]
                $^b$ Department of Particle Physics, Weizmann Institute of Science,\\
                     Rehovot 76 100, Israel\\[4mm]
                $^c$ Department of Physics, Indian Institute of Technology, Powai,\\
                     Mumbai 400 076, India
                \end{center}}

\vspace*{0.5cm}
\date{today}
%
%
%
%
\begin{abstract}
We study the problem of determination of the sign of $\da$, or the
neutrino mass hierarchy, through observations of atmospheric neutrinos in future detectors.
We consider two proposed detector types :
 (a) Megaton sized water \chr detectors, which can measure the
event rates of $\nu_\mu + \bar{\nu}_\mu$ and $\nu_e + \bar{\nu}_e$
and  (b) 100 kton sized magnetized iron detectors, which can measure the
event rates of $\nu_\mu$ and $\bar{\nu}_\mu$.
For energies and path-lengths relevant to atmospheric neutrinos,
these rates obtain  significant matter  contributions from
$\pmue$, $\pmumu$ and $\pee$, leading to an appreciable
sensitivity to the hierarchy.
We do a binned $\chi^2$ analysis of simulated data in these two types of
detectors which includes the effect of smearing in neutrino energy and
direction and incorporates detector efficiencies and
relevant statistical, theoretical and systematic errors.
We also marginalize the $\chi^2$  over the  allowed ranges
of neutrino parameters in order to accurately  account for their uncertainties.
Finally, we  compare the performance of both types of
detectors  vis a vis the hierarchy determination.
\end{abstract}
%
%
%

\newpage
\renewcommand{\thefootnote}{\arabic{footnote}} 
\setcounter{footnote}{0}
%
%
%
\section{Introduction}
Our understanding of  neutrino physics has progressed impressively
over the last decade owing to outstanding results from
solar, atmospheric, reactor and accelerator based neutrino oscillation
experiments. For three neutrino flavors, the picture of neutrino masses
and mixings emerging from these results is:
\bi
\item There are two independent mass squared differences~\footnote{We use the convention ${\mathrm{\dji \equiv m_j^2 - m_i^2.
}}$} given by
$|\da| \sim 10 ^{-3}$ eV$^2$ and $\ds \sim 10^{-5}$ eV$^2$
%
;
\item Of the three mixing angles, two ($\theta_{12}$ and $\theta_{23}$) are
large, while the third mixing angle ($\theta_{13}$)
 is small and constrained only by an upper bound.
\ei
In \Tab~\ref{param}, we summarize the present bounds (the best-fit values and
3$\sigma$ ranges) of these parameters.



\begin{table}[htb]
\begin{center}
\begin{tabular}{ c | c |  c }
\hline\hline
& &\\
\hspace{0.5cm}{\sf {Parameter}}\hspace{0.5cm} & \hspace{0.5cm}{\sf
{Best-fit value}}\hspace{0.5cm} &
\hspace{0.5cm}{\sf {$3\sigma$
range}} \hspace{0.5cm}
\\
&&\\
\hline
$\ds [10^{-5} {\mathrm{eV}}^2]$ & 7.9 &   7.1 -- 8.9
\\
$|\da| [10^{-3} {\mathrm{eV}}^2]$ & 2.5 &   1.9 -- 3.2
\\
$\sin^2 \theta_{12} $ & 0.3 &   0.24 -- 0.40
\\
$\sin^2 \theta_{23}$ & 0.5 &  0.34 -- 0.68
\\
$\sin^2  \theta_{13}$  & 0.0 & $\le$ 0.041
\\
\hline
\hline
\end{tabular}
\mycaption{{Best-fit values  and $3\sigma$ allowed
intervals for three-flavour
neutrino oscillation parameters from combined analysis of global
data including solar, atmospheric, reactor (\KamLAND and \CHOOZd) and
accelerator (\KTOK and \MINOSd) experiments~\cite{Schwetz:2006dh}.
}} \label{param}
\end{center}
\end{table}
%
These results have also delineated  the main goals of neutrino
oscillation research over the next decade, which
can be summarized as follows :
\bi
\item
Improved  precision  on  the mass squared differences
($\ds$ and $\da$) and the mixing angles ($\theta_{23}$ and $\theta_{12}$)~;
\item
Determination of $\theta_{13}$, especially ascertaining if its value is non-zero~;
\item
Determination of $\signda$ or the  hierarchy of neutrino masses~;
\item
The presence, or otherwise, absence of CP violation in the leptonic sector and
the value of the  phase $\dcp$.
\ei

The experimental realization of the above goals is a complex task.
Hence, a large number of neutrino experiments are being constructed
or being planned to work towards achieving these aims.

Our focus in this work is to bring out the potential of megaton
water \chr detectors (\eg~\HKd~\cite{Nakamura:2000tp,Itow:2001ee},
\UNOd~\cite{Jung:1999jq} or \MEMPHYSd~\cite{deBellefon:2006vq})
and magnetized iron detectors (\eg~\INOd~\cite{Athar:2006yb}) vis
a vis the goal of determining the neutrino mass hierarchy. Both of
these detector-types  plan to use atmospheric neutrinos as one of
their primary sources. Each category of  detector has a special
ability, in addition to large statistics resulting from its
massive size. In the case of water \chr detectors, one can detect
both muons and electrons, whereas in the case of magnetized iron
detectors, it is possible to study muons and anti-muons separately
by charge discrimination. These advantages, together with their
size, and the wide ranges in energy and path-length made available
by atmospheric neutrinos, make these detectors powerful tools for
studying $\signda$.

Solar neutrino data require $\ds$ to be positive. However, data from
atmospheric neutrino experiments as well as  accelerator neutrino
experiments (\KTOK and \MINOSd) constrain only the magnitude of
$\da$ but not its sign.
If $\signda > 0$, then we have the  mass pattern, ${\mathrm{m_3 \gg m_2 \gg m_1}}$, which is
similar to that of  the charged leptons. This is called the  normal hierarchy (NH).
If $\signda < 0$, then the mass pattern is ${\mathrm{m_2 \geq m_1 \gg m_3}}$.
This is called the  inverted hierarchy (IH).
These statements, of course, are meaningful provided a discernible and
non-trivial hierarchy exists
among the neutrino mass eigenstates, or, in other words,
the lightest neutrino state is almost massless.
If future experiments
on tritium beta decay or neutrinoless double beta decay show that
the absolute scale of neutrino masses is, for instance,  $\gsim 0.3$ eV, then
all three neutrino masses must be  nearly degenerate.
However, one may still ask the question if $\signda$ is positive or
negative.

Attempts to construct unified theories beyond the Standard Model
depend crucially on the hierarchy. In fact, one way to
classify families of unification models is via the hierarchy they
assume as input. It is thus an important ingredient in
our quest for a unified theory, and its determination would in
essence eliminate or strongly disfavour large classes of such theories
and considerably narrow the focus of this search.

Generally speaking, determination of the mass ordering in
oscillation experiments requires
the observation of measurably large matter effects (\ie~long
baselines) and a not too small value for $\stsmall$ $ (\gsim~0.05)$.
This limits the sensitivity of presently planned  superbeam experiments to
hierarchy determination as the baselines involved are moderate
($\lsim$ 800 km) and significant matter effects do not
develop.
Also at these baselines, for the $\pmue$ channel which these experiments
use, the oscillation
probability depends jointly on  three unknown parameters:  $\signda$,
the phase, $\dcp$ and the mixing angle, $\theta_{13}$.
This results in a $(\dcp,\signda)$ degeneracy
where acceptable solutions corresponding to the two signs of $\da$ can appear
at different values of $\dcp$ and $\theta_{13}$~\cite{Minakata:2001qm}~\footnote{The $\dcp$ dependence
tends to decrease for
longer baselines, and in fact  vanishes at (and around) the so called
``magic baseline" ($\sim 7000$ km)~
\cite{Barger:2001yr,Huber:2003ak,Smirnov:2006sm,Gandhi:2006gu,Minakata:2006am}, reappearing subsequently at longer baselines.}.
To overcome this problem, the synergistic use of two
experiments~\cite{Barger:2002xk,Burguet-Castell:2002qx,Wang:2001ys,Whisnant:2002fx,Huber:2003pm,Donini:2004hu}
or  two detectors at different baselines\cite{Mena:2005ri,Ishitsuka:2005qi} has been suggested.
Among the combinations studied, the planned superbeam experiments \THKd~\cite{Hayato:2005hq,Kudenko:2005rr}
and the NuMI Off-axis experiment \NOVAd~\cite{Ayres:2004js} may be able to infer the
neutrino mass hierarchy~\cite{Huber:2004ug,Minakata:2003ca,Mena:2004sa,Mena:2005sa}.

In this paper, in order to ascertain the neutrino mass hierarchy,
we focus on the appreciable  resonant matter effects encountered
by atmospheric neutrinos with energies between 2$-$10 GeV which
traverse distances between 4000$-$12500 km inside the  earth.
Until very long baseline experiments using $\beta$-beams or
neutrino factories are built atmospheric neutrinos  are the only
source that permit us to exploit these spectacular
effects~\cite{Agarwalla:2006vf,Gandhi:2006gu,Akhmedov:2005yj,Cervera:2000kp,Albright:2000xi,Freund:2001ui,Choubey:2006jk}.
Moreover, for these energy and pathlength ranges, it turns out
that the effects of $\dcp$ degeneracy are not very
significant~\cite{Gandhi:2004bj}. Therefore, lately, the subject
of hierarchy determination using atmospheric neutrinos has
received considerable
attention~\cite{Banuls:2001zn,Bernabeu:2001xn,Indumathi:2004kd,Gandhi:2004bj,
Petcov:2005rv,Samanta:2006sj}. Recent work has also emphasized the
degeneracy resolving power of atmospheric neutrino
data~\cite{Huber:2005ep,Campagne:2006yx, Donini:2007qt} in the
context of water \chr and magnetized iron  detectors respectively.
In particular, the issue of the  resolution of
$(\theta_{23},\pi/2-\theta_{23})$ degeneracy in magnetized iron
detectors has been  considered
in~\cite{Choubey:2005zy,Indumathi:2006gr}.

In our earlier work on this subject~\cite{Gandhi:2004bj,Gandhi:2005wa}
we had used the total event count rates (in the energy range 2$-$10 GeV and pathlength range
4000$-$10000 km) to examine the determination of the neutrino mass hierarchy for iron magnetized calorimeters
and water \chr detectors respectively.
Since we used total event rates, the effects of energy and angular smearing were not very important
in those analyses.
We  had also made the simplistic assumption that the values of the neutrino
parameters are known exactly. In this paper, we relax this
assumption and take into account the uncertainties in the determination
of neutrino oscillation parameters.
Under these circumstances, the total event rates can no longer give good
sensitivity and therefore one has to do a binned spectrum analysis
including  appropriate angular  and energy smearing.
In what follows, we do a detailed study of
the effect of these refinements on the hierarchy sensitivity
for both water \chr and iron calorimeter detectors.

The hierarchy sensitivity of iron calorimeter detectors was studied
in~\cite{Petcov:2005rv} under the approximation that $\ds = 0$.
In our numerical work we take $\ds$ to be non-zero because, in principle, it
becomes important in the limit  of small $\theta_{13}$.
However, the marginalization over $|\da|$ washes out the
hierarchy sensitivity due to non-zero $\ds$ in this limit.
Allowing a non-zero value for $\ds$ allows us to
study the variation of $\chi^2$ with $\dcp$.
We also compare and contrast the performance of a magnetized iron calorimeter
detector with a water \chr detector in
discriminating between the two hierarchies.

In the next section, we discuss and summarize the matter sensitivities
of various oscillation probabilities that are relevant
to the atmospheric neutrino signal and give analytic expressions for
maximum hierarchy sensitivity.
\Sec~3 describes the prototype  detectors we have used and the details
of the  statistical analysis on which our results are based. We also
include a description of our treatment of  energy and angular smearing
and our choices of the number of bins for each of these variables.
\Sec~4 is devoted to a description of
our numerical procedure on the $\chi^2$ sensitivity
to the hierarchy. \Sec~5 summarizes our results and conclusions.
%
%
%
%

%
%
%

\section{Earth-matter effects in atmospheric neutrino oscillation probabilities}

\begin{figure}
\centering
\begin{tabular}{c@{\quad \quad \quad \quad \quad \quad}c}
\includegraphics[height=13cm,width=7.2cm]{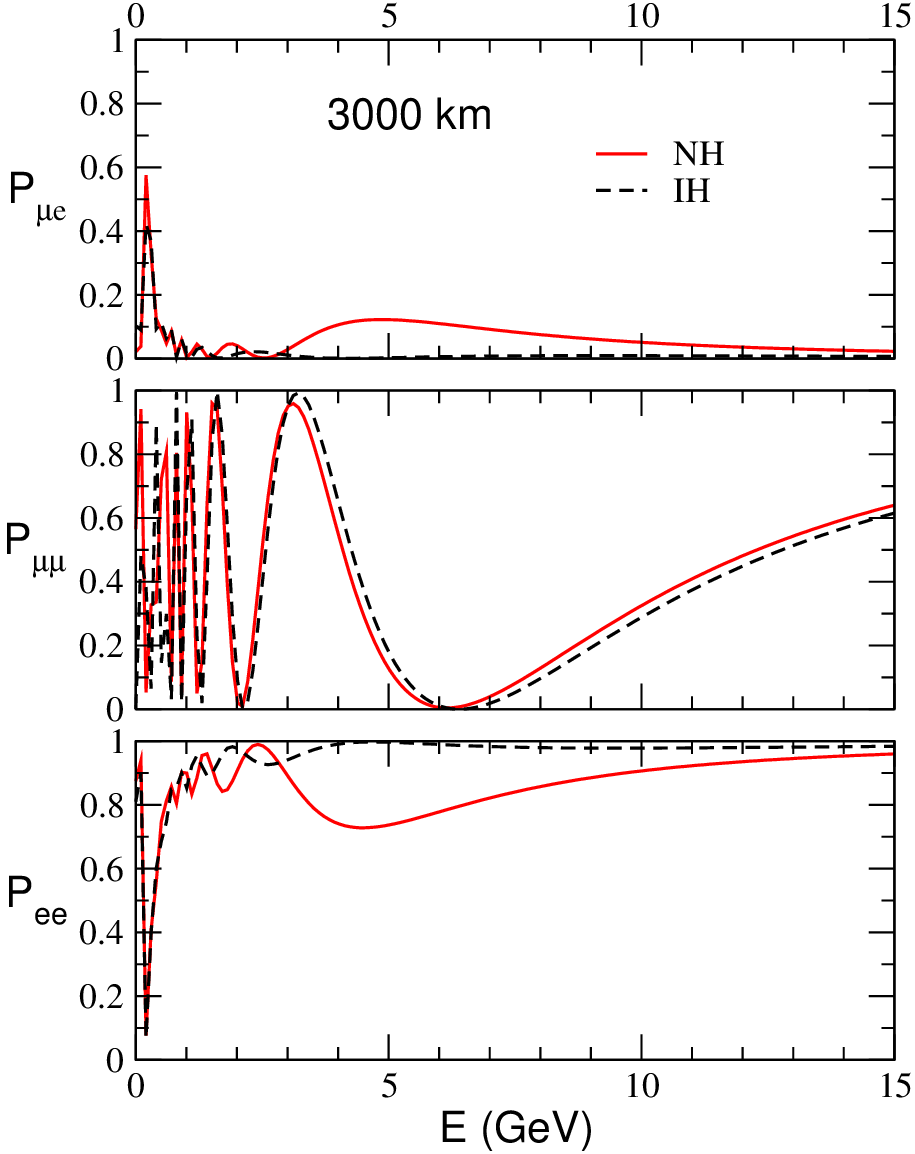}
&
\includegraphics[height=13cm,width=7.2cm]{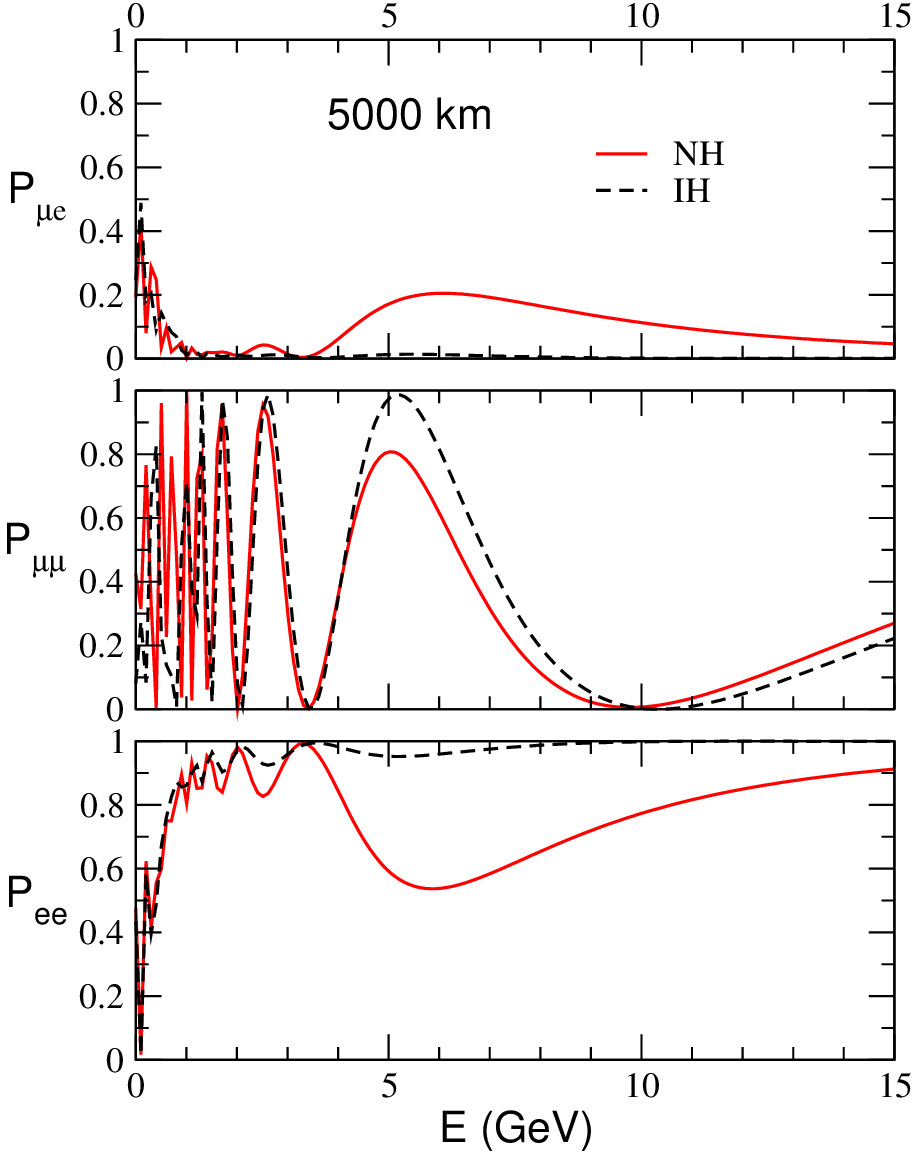}
\end{tabular}
\mycaption{The probabilities $\pmue,\pmumu$ and $\pee$ in matter plotted versus
neutrino energy for two baselines, 3000 km (left panel) and 5000 km (right panel).
The solid curves correspond to NH ($\da > 0$), while the dashed curves are for IH ($\da < 0$).
Here, we use the neutrino parameter values listed in \eq~\ref{paramvalues}.}
\label{fig1}
\end{figure}

\begin{figure}
\centering
\begin{tabular}{c@{\quad \quad \quad \quad \quad \quad}c}
\includegraphics[height=13cm,width=7.2cm]{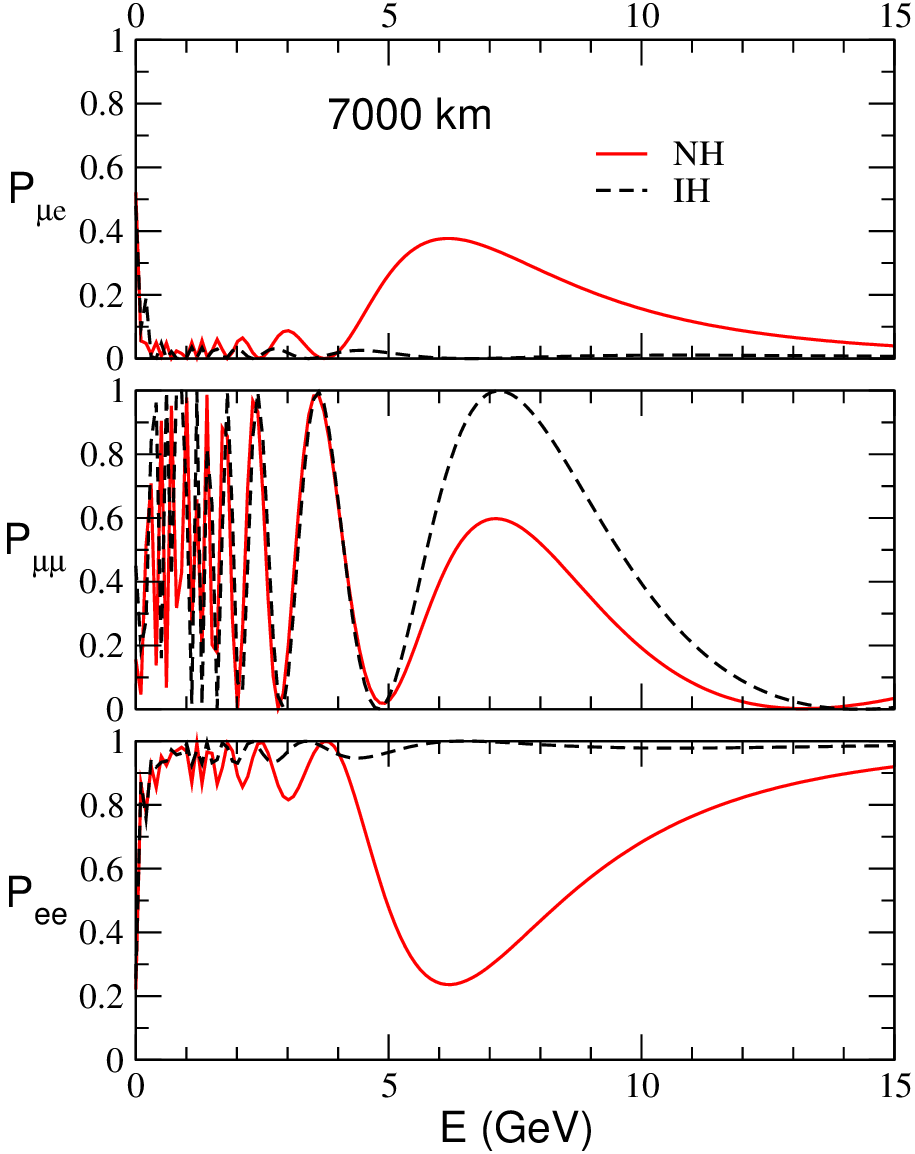}
&
\includegraphics[height=13cm,width=7.2cm]{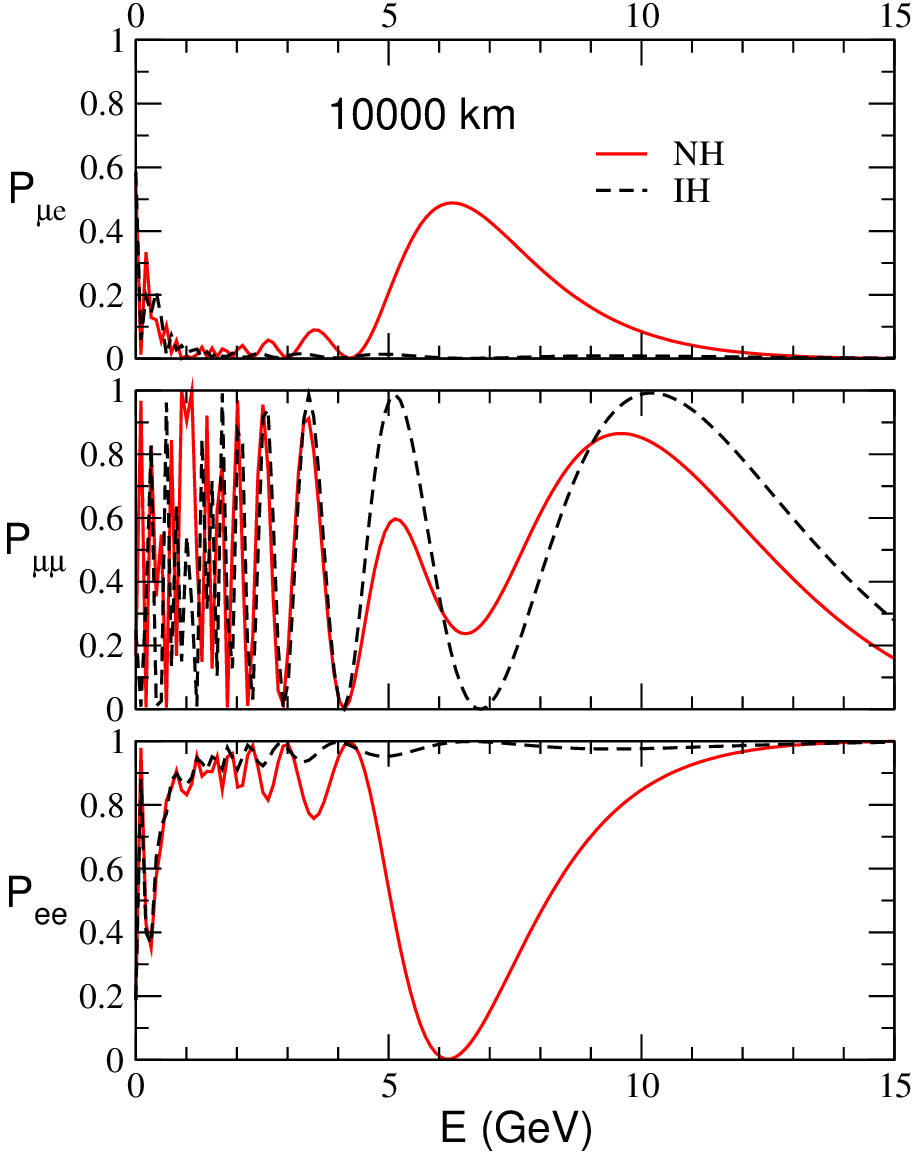}
\end{tabular}
\mycaption{The probabilities $\pmue,\pmumu$ and $\pee$ in matter
plotted versus neutrino energy for two baselines, 7000 km (left
panel) and 10000 km (right panel). The solid curves correspond to
NH ($\da > 0$), while the dashed curves are for IH ($\da < 0$).
Here, we use the neutrino parameter values listed in
\eq~\ref{paramvalues}.}
\label{fig2}
\end{figure}

In calculating atmospheric electron and muon event rates, the neutrino oscillation probabilities
corresponding to the disappearance channels
$\pmumu$ and  $\pee$
and appearance channels
$\pmue$ and $\pemu$
are of direct relevance~\footnote{${\mathrm{P_{\alpha\beta}}}$ denotes the
probability for transition from $\nu_{\alpha} \to \nu_{\beta}$.
}.
For the long path-lengths under consideration here,
we need to explicitly take into account the
varying density profile of the earth, and for this purpose we
 use the  Preliminary Reference Earth Model (PREM)~\cite{prem}.
In \Figs~{\ref{fig1}} and {\ref{fig2}}, we plot the three
probabilities $\pmue$, $\pmumu$ and $\pee$ as a function of energy for
both normal and inverted hierarchies for four different path-lengths in the
range 3000$-$10000 km.
These curves result from  numerically solving the full three flavour
neutrino propagation equation through the earth. In obtaining
them we have used the following values for neutrino parameters:
\bea
\vert \da \vert = 2.5 \times 10^{-3}~ {\mathrm{eV^2}}~,~~~~~\ds = 8.0 \times 10^{-5}~ {\mathrm{eV^2}}, \nonumber\\
\sin^2 \theta_{12} = 0.31~,~~~~~\sin^2 2\theta_{23} = 1.0~,~~~~~\sin^2 2 \theta_{13} = 0.1
\label{paramvalues}
\eea

%
%

%
%

\Figs~{\ref{fig1}} and {\ref{fig2}} demonstrate the following qualitative features :
\bi
\item The differences between the probability values for the two types of hierarchy in  all the
channels are maximized  in the energy range 4$-$10 GeV.
\item Matter effects in $\pmue$ in case of a NH manifest themselves by a {\it{rise}}
over the corresponding value for an IH over the range of energies and baselines
considered in these figures.
\item On the other hand, matter effects in $\pmumu$  for a NH
manifest themselves as a {\it decrease} over the IH value in certain energy and baseline
ranges and an {\it increase} in others (\eg~for baselines around 10000 km).
\item The value of $\pee$ for a NH consistently shows a {\it{fall}}
compared  to the value with an  IH for all four baselines
in the energy range 2$-$10 GeV.  This fall can be as high as 100\% (\eg~for baselines around 10000 km).
\ei

We discuss these features below, using the one mass scale dominant (OMSD)
approximation, which amounts to neglecting the smaller mass-squared difference $\ds$
in comparison to $\da$.
The condition for this approximation to be valid is
\be
{\mathrm{\frac{\ds L}{E}}} << 1
\label{omsd}
\ee
Alternatively, this corresponds to ${\mathrm{{L/E} << 10^4}}$ km/GeV. For
multi-GeV neutrinos, this condition is violated only for a small fraction
of events with E $\simeq$ 1 GeV and ${\mathrm {L \geq 10^4}}$ km. Hence,
effectively \eq~\ref{omsd} is valid for most of the energies and pathlengths under consideration here.
Additionally, the OMSD approximation is not valid for very small $\theta_{13}$,
since the terms containing the small parameter $\ds$ can be dropped only
if they are small compared to the leading order terms containing $\theta_{13}$
~\footnote{Expressions of probabilities containing corrections of the order of $\ds/\da$ can for
instance be found in~\cite{Freund:2001pn,Akhmedov:2004ny,Kimura:2002hb,Takamura:2005df}.}.

Making this approximation and assuming constant matter density, the probabilities
$\pmue$, $\pee$ and $\pmumu$ in matter are given by
%

\begin{align}
{\mathrm{P^{m}_{\mu e}}} =
{\mathrm{
\sin^2 \theta_{23} \; \sin^2 2 \theta^m_{13} \; \sin^2\left[1.27 ~\dam ~\frac{L}{E} \right]
}}
\label{eq:pemumat}
\end{align}
The probability  for the time-reversed transition $\pemum$ is same as $\pmuem$ with the replacement
$\dcp\to-\dcp$ for a symmetric matter density profile like the PREM profile.
The OMSD analytical expressions are insensitive to $\dcp$, hence, $\pemum = \pmuem$ in this limit.
In general, the two probabilities will be equal only when $\dcp = 0$.

\begin{align}
{\mathrm{P^{m}_{e e}}} =
{\mathrm{
1 - \sin^2 2 \theta^m_{13}\; \sin^2\left[1.27 ~\dam~\frac{L}{E} \right]
}}
\label{eq:peemat}
\end{align}

\bea
{\mathrm{P^{m}_{\mu \mu} }} &=&
{\mathrm{1 - \cos^2 \theta^m_{13} \; {\mathrm{\sin^2 2 \theta_{23}}}
\;
\sin^2\left[1.27 \;\left(\frac{\da + A + \dam}{2}\right) \;\frac{L}{E} \right]}}
\nonumber \\
&& ~-~
{\mathrm{
\sin^2 \theta^m_{13}\; {\mathrm{\sin^2 2 \theta_{23}}}
\;
\sin^2\left[1.27 \;\left(\frac{\da + A - \dam}{2}\right) \;\frac{L}{E}
\right]}}
\nonumber \\
&& ~-~
{\mathrm{\sin^4 \theta_{23}}} \;
{\mathrm { \sin^2 2\theta^m_{13} \;
\sin^2 \left[1.27\; \dam  \;\frac{L}{E} \right]
}}
\label{eq:pmumumat1}
\eea

 The  mass squared difference ${\mathrm{\dam}}$ and mixing angle
${\mathrm {\stsmallm}}$ in matter are related to their vacuum values by
\bea
{\mathrm{\dam}} &=&
{\mathrm{
\sqrt{(\da \cos 2 \theta_{13} - A)^2 +
(\da \sin 2 \theta_{13})^2} }}
\nn \\
\nn \\
{\mathrm {\sin 2 \theta^m_{13} }}
&=& \frac{\mathrm{\da \sin 2 \theta_{13}}}
{{\mathrm{\sqrt{(\da \cos 2 \theta_{13} - A)^2 +(\da \sin 2 \theta_{13})^2} }}}
\label{eq:dm31}
\eea
where
$${\mathrm{A = 2\;\sqrt{2}\;G_F\;n_e\;E = 2 \times 0.76 \times 10^{-4} \times Y_e
\;\left[\frac{\rho}{g/cc}\right]
\;\left[\frac{E}{GeV}\right]\;eV^2 }}$$ is the MSW matter
potential~\cite{Mikheyev:1989dy,Wolfenstein:1977ue} which depends
on the Fermi coupling constant, $\rm{G_F}$, the number density of
electrons, ${\mathrm{n_e}}$ and energy of the neutrinos, E.
${\mathrm{Y_e}}$ is the fraction of electrons, which is $\simeq
0.5$ for earth matter and $\rho$ is matter density inside earth.
The anti-neutrino probabilities can be written down by making the
replacement A $\to$ -A in above equations.

We note that ${\mathrm{(\Delta m_{31}^2)^m }}$ and ${\mathrm{\sin 2 \theta^m_{13}}}$ can assume very different
values for a NH and an IH, leading to hierarchy dependant differences in all three probabilities given above.
Furthermore, large changes in the values of these probabilities arise not just due
  to the resonance occurring  when $\stsmallm \to 1$, $\ie~{\mathrm{A = \da \cos 2 \theta_{13}}}$, but also due to the matter dependence of the oscillatory L/E terms contained
in them. Thus the full effect results from the product of the two types of terms assuming values significantly different from those in vacuum
\cite{Gandhi:2004bj,Akhmedov:1988kd,Akhmedov:2005yj}.

The matter resonance, which occurs for neutrinos when the
hierarchy is normal and for anti-neutrinos when it is inverted,
requires \be {\mathrm {E (GeV)}} \simeq {\mathrm {E_{res} (GeV) }}
= \left[\frac{\mathrm{1}}{\mathrm{2 \times 0.76 \times 10^{-4}
Y_e}}\right]
\left[\frac{{\mathrm{|\da|}}}{\mathrm{eV^2}}\right]
\left[\frac{\mathrm{g/cc}}{\rho}\right]
\label{eq:eres}
\ee
If we use an average density calculated using PREM profile of earth,
then we get the resonant energy at various baselines, as listed, in \tab~\ref{eresvalues}.


%
%
\begin{table}[htb]
\begin{center}
\begin{tabular}{ c | c | c  }
\hline \hline
&& \\
\hspace{0.5cm}{\sf {L (km)}}\hspace{0.5cm} & \hspace{0.5cm}${\sf{\rho_{avg} (g/cc)}}$\hspace{0.5cm} & \hspace{0.5cm}$
{\sf E_{res}}$ \sf{(GeV)}\hspace{0.5cm}
\\
 && \\
        \hline
 3000 &  3.32  &  9.4 \\
 5000  &   3.59 &  8.7 \\
 7000   & 4.15 & 7.5 \\
 10000  & 4.76 &  6.6 \\
\hline \hline
\end{tabular}
\mycaption{Values of ${\mathrm{E_{res}}}$ at 13 resonance in
case of OMSD are listed
as a function of baseline and the average density ${\mathrm{\rho_{avg}}}$.
We have used $|\da|=2.5 \times 10^{-3}$ \evsq
and $\sin^2 2\theta_{13}=0.1$ in \eq~\ref{eq:eres}.}
\label{eresvalues}
\end{center}
\end{table}

Clearly, hierarchy sensitivity is enhanced  when the difference
\be
{\mathrm{{\Delta P}_{\alpha \beta}}}={\mathrm{P^{m}_{\alpha \beta}(NH)}}-{\mathrm{P^{m}_{\alpha \beta}(IH)}},
\ee
 is large~\footnote{Here $\alpha, \beta$ may be e or $\mu$. ${\mathrm{P^{m}_{\alpha \beta}(IH)}}$
is computed in each case by reversing the sign of $\da$ in the expression
for ${\mathrm{P^{m}_{\alpha \beta}}}$.}.
It is useful to use the OMSD expressions given above to examine the
conditions required for this to occur,
and to co-relate them with the large differences between the NH and IH curves
visible in \Figs~{\ref{fig1}} and {\ref{fig2}} (which, as mentioned earlier,
have been obtained using the full three flavour
evolution equation and the PREM profile).
We proceed to do this below for each of the three probabilities in turn.

\subsection{$\pmue$}

From \eq~\ref{eq:pemumat}, we can see that
this probability reaches its maximum value of $\sin^2\theta_{23}$
when the
resonance  condition, ${\mathrm{\sin ^2 2\theta_{13}^m}} = 1$
and the condition
${\mathrm{\sin^2 (1.27 (\Delta m_{31}^2)^m {\mathrm L}/{\mathrm E}) = 1}}$ are
simultaneously satisfied.
The second condition yields an  energy (henceforth referred to as the {\bf{matter peak energy}}) given by
\be
{\mathrm{
E^{m}_{peak} = \frac{1.27}{(2p+1)\;\pi/2} ~{\dam \;L}}}~;\, p=0,1,2,\ldots
\ee
Thus, the condition for obtaining a maxima in $\pmue$ can be expressed as
${\mathrm{E_{res} = E^{{m}}_{{peak}}}}$. This determines the distance for
maximum matter effect via
\be
{\mathrm{[\rho \, L]^{peak}_{\mu e} =
{5.18 \times 10^3} \; \pi ~\frac{(2p+1)}{\tan 2\theta_{13}} ~{\rm km ~g/cc}}}
\label{eq:eecondtn}
\ee
The above equation is independent of $\da$ but depends on the vacuum value of $\theta_{13}$.
Note, from the above equation  however, that for a given resonant energy,
whether or not the oscillatory term in L/E can become maximal in a meaningful way depends on:
\benu
\item[(i)] p being low enough and $\stsmall$ large enough that the resultant L is less than the earth's diameter, ${\rm{L_D}}$ = 12742 km. For $\stsmall$ =  0.1, the only allowed value of p is p=0, and this gives L=10200 km in conjunction with ${\mathrm{\rho_{avg} = 4.8}}$ g/cc.
\item[(ii)] $\stsmall$ not being so small that the OMSD approximation, which we use for the discussion in this section, is invalidated.
\eenu
For our discussion which follows below leading to \eq~\ref{eq:delpmue} and \eq~\ref{eq:delpmumu}, the first restriction on $\stsmall$ limits $\stsmall\gsim 0.04$~\footnote{ $\stsmall\gsim 0.04$ is obtained by substituting L = ${\rm{L_D}}$ = 12742 km
and $\rho \sim 7$ g/cc. This value is a conservative estimate considering the fact that
 density at the center of earth is not very well-known and that at such large distances, the constant density approximation is not very good.
}.

Under these conditions, for a particular value of $\theta_{13}$, one may write,
\bea
{\mathrm{\Delta P_{\mu e}^{max}}}   & = &
{\mathrm{ \sin^2\theta_{23} \Bigg[1 - \frac{\sin^2 2\theta_{13}}{4-3\sin^2 2\theta_{13}}~
\sin^2\left[ (2p+1)\;\frac{\pi}{2}\; \frac{1}{\sin 2\theta_{13}} \;{\sqrt{4-3\sin^2 2\theta_{13}}} \right]\Bigg]}}
\label{eq:delpmue}
\eea

Numerically, this reduces to ${\mathrm{{\Delta P_{\mu e}^{max}}\simeq 0.5}}$ for $\theta_{23} = \pi/4$
and $\stsmall =  0.1$. Comparing with the top panel of the right-hand set of plots  in \Fig~\ref{fig2}, we see that
 both the baseline
value and  ${\mathrm{\Delta P_{\mu e}^{max}}}$ are in very good agreement with these estimates.
For smaller baselines although the resonance condition is achieved the
oscillatory term remains $<1$ and thus the rise in $\pmuem$ in matter for NH
is lower. However, since this is an appearance channel even with a moderate
increase in probability it is possible to do interesting physics if the
backgrounds are well understood.

\subsection{$\pee$}


\noindent From \eq~\ref{eq:peemat} it is clear that
the condition for obtaining maximum matter effect in $\peem$ is the same as that for
$\pemum$, \ie, ${\mathrm{E_{res} = E^{{m}}_{{peak}}}}$. This gives
\bea
{\mathrm{\Delta P_{e e}^{max}}}
&=&
{\mathrm{\Bigg[-1 + \frac{\sin^2 2\theta_{13}}{4-3\sin^2 2\theta_{13}}~
\sin^2\left[(2p+1)\;\frac{\pi}{2}\; \frac{1}{\sin 2\theta_{13}}\;\sqrt{4-3\sin^2 2\theta_{13}}\right]
\Bigg]
}}
\label{eq:delpee}
\eea
This quantity is maximized when $\peem$ (NH) is at its minimum value of 0
and $\peem$ (IH) is at its maximum value, which is close to 1.
For $\stsmall = 0.1.$ and the earlier baseline value of $\approx 10000$ km, this reduces
numerically to $\simeq -0.98$, which is manifest in the bottom
panel of the right-hand set of curves in \Fig~\ref{fig2}~\cite{Agarwalla:2006gz}.

\subsection{$\pmumu$}


The muon disappearance probability is a somewhat more complicated function  as can be seen from \eq~\ref{eq:pmumumat1}.
In order to understand where and why maximal hierarchy effects arise, we
first note from \Figs~\ref{fig1} (central right-hand panel)
and \ref{fig2} (central panels in both left-hand and right-hand sets)  that
large hierarchy sensitivity occurs  when the neutrino energy
is close to ${\mathrm {E_{res}}}$ (cf \Tab~2) {\it and} in the vicinity
of a peak or dip in $\pmumum$(IH).
Since the matter effect in neutrinos is negligible for IH, these curves closely follow the
vacuum probability, and in particular have peaks and dips at the same locations.

Noting that the  vacuum peak of $\pmumu$ occurs when
\bea
1.27~\frac{\da \mathrm L}{\mathrm {E^v_{peak}}}
= {\rm p} \;\pi
\label{eq:pmumuvacpeak}
\eea
and, setting ${\mathrm{E^v_{peak} = E_{res}}}$, we obtain
\bea
{\mathrm{[\rho L]_{\mu \mu}^{peak} \simeq {p\,\pi\,\times10^{4}
\, \times \, \cos 2\theta_{13}}~{km~g/cc}}}
\label{eq:mumucondtn}
\eea

For p$=$1, \ie, when the resonance is near the first vacuum peak
this leads to L$\simeq$ 7000 km (taking the average density to be 4.15 g/cc),
which is borne out by the middle panel of the
left column of \Fig~\ref{fig2}. The magnitude of this fall can be estimated by~\footnote{In order to simplify this expression, we have set $\cos^2 2\theta_{13}=1$.}
\bea
{\mathrm{\Delta P_{\mu \mu}^{max,peak}}}  & \simeq &
\Bigg[
{\mathrm {-\sin^2 \left[p\;\frac{\pi}{2} \; \sin2\theta_{13}\right]}}
- {\mathrm {\frac{1}{4}\; \sin^2\Big[p\;\pi\;
\sin2\theta_{13} \Big]}} \nonumber \\
&+&
{\mathrm{\sin^2\left[p\;\frac{\pi}{2}\; \sqrt{4-3\sin^2 2\theta_{13}}\right]}} \nonumber \\
&+&
{\mathrm{\frac{1}{4}~\frac{\sin^2 2\theta_{13}}{4-3\sin^2 2\theta_{13}} ~
\sin^2\left[p\;\pi\;\sqrt{4-3\sin^2 2\theta_{13}}\right]}}
\Bigg]
\label{eq:delpmumu}
\eea
Numerically, this gives a  drop in $\pmumum$ (NH) relative to $\pmumum$ (IH) of $\simeq -0.4$,
which is borne out by central panel in \Fig~\ref{fig2} near the region of peak in IH curve.

The right column of \Fig~\ref{fig2} also shows that at 10000 km
there is a {\it rise}
in  $\pmumu$(NH) relative to $\pmumu$(IH) near a
dip in the IH probability at $\sim$ 6 GeV.
The condition for dip in $\pmumu$ in vacuum is
\bea
1.27~\frac{\da \mathrm L}{\mathrm {E^{v}_{dip}}}
= {\mathrm{(2p+1) \;\frac{\pi}{2}}}
\label{eq:pmumuvacdip}
\eea
This gives the position of the second vacuum dip (\ie~when ${\mathrm{p=1)}}$ for ${\mathrm{L=10000}}$ km to be around $\sim$ 6.7 GeV.
\tab~\ref{eresvalues} shows that at 10000 km  the resonance energy is $\sim$ 6.6 GeV.
Thus, we have the condition ${\mathrm{{E_{res}} \approx E^{v}_{dip}}}$
satisfied here for p$=$1.
Using this, one can similarly estimate the magnitude of the corresponding rise, and find it to be in good agreement
with the figure, \ie~about $22\%$ for ${\mathrm{p=1}}$~\footnote{For ${\mathrm{p=0}}$,
the value of ${\mathrm{\Delta P_{\mu \mu}}}$ is  small since matter effects are not appreciable at the associated baseline
of 5000 km.}.

It is interesting to note from \eq~\ref{eq:delpmue} and from
\Fig~\ref{fig2} that at this baseline there is a 50\% rise in
$\pmue$ probability over IH.  Since $\Delta \pmutau = -(\Delta
\pmue + \Delta \pmumu)$ this implies a 70\% matter induced
decrease in the $\pmutau$ probability, which is a transition
probability between two species of neutrinos for which matter
effect occurs only via neutral current interactions and is
identical to both. This dramatic matter driven decrease is a
genuine three flavour effect and was first pointed out
in~\cite{Gandhi:2004md}.

To summarize, we have discussed the important features of 3
generation oscillation probabilities in matter corresponding to
appearance and disappearance channels. In order to do that, we
have (in this section only) worked in the OMSD approximation,
which allows the use of tractable analytic expressions. We have
emphasized those features which percolate into the disappearance
probability for electron and muon events and modify the expected
atmospheric neutrino signal in water \chr and magnetized iron
detectors in a hierarchy dependent manner, and examined the cases
where these effects are maximized. However, the full effects are
spread
 over a wide band of energies and baselines (4$-$12 GeV and 3000$-$12000 km~\footnote{Beyond 10500 km, the neutrinos start traversing  the core, causing the onset of  mantle-core interference effects~\cite{Petcov:1998su,Bernabeu:2001xn,Akhmedov:1998xq,Akhmedov:1998ui}.
Our full numerical calculations incorporate the difference between NH and IH probabilities due to these effects.}
 respectively).
Inspite of the fact that the effects vary in significance over these ranges for each of the probabilities,
cumulatively they provide a powerful discriminator of the hierarchy.
We exploit this fact by doing a bin-by-bin $\chi^2$ analysis of simulated data to
determine the potential for hierarchy determination.

%
%
%
%
%
\section{Detector Characteristics}

In this section we describe the details of the two types of
detectors that we have considered for our study.


\subsection{Water \chr Detectors}

We consider a prototype megaton Water \chr detector based on the proposed  Hyper-Kamiokande (\HKd)
detector. \HK will essentially be a scaled up
version of the  Super-Kamiokande (\SKd) detector with
the  total volume increased to 1 Mt and a fiducial volume of 545 Kt.
 R\&D initiatives are in progress currently to study the non-trivial physics and engineering
 issues which
arise due to this scaling up in size~\cite{Rubbia:2004nf}.

Since water \chr detectors can discriminate between  muon and
electron events, the full atmospheric neutrino spectrum can be
studied. They are, however, insensitive to lepton charge, thus
neutrino and anti-neutrino events  must be added  together. In the
following, we add the events of $\nu_\mu$, whose charged current
(CC) interactions produce a $\mu^-$, and those of $\bar{\nu}_\mu$,
whose CC interactions produce a $\mu^+$, and call the sum to be
muon events. Similarly, we add the CC events of ${\mathrm{\nu_e}}$
and ${\mathrm{\bar{\nu}_e}}$ and call the sum to be electron
events.

In this case the hierarchy sensitivity is determined by the difference
in the total number of events. For instance for muon events this quantity is
\be
{\mathrm{
\Delta N  =
(N_{\mu^-}^{NH} + N_{\mu^+}^{NH}) - (N_{\mu^-}^{IH}+
 N_{\mu^+}^{IH})
 =  (N_{\mu^-}^{NH} -  N_{\mu^-}^{IH}) +
    (N_{\mu^+}^{NH} -  N_{\mu^+}^{IH})
= \Delta N_{\mu^-} + \Delta N_{\mu^+}
}}
\label{eq:Deltan}
\ee
If the hierarchy is normal, matter effects induce large changes in
neutrino appearance and disappearance probabilities and hence in
${\mathrm{N_{\mu^{-}}^{NH}}}$, while leaving anti-neutrino
probabilities  and hence ${\mathrm{N_{\mu^{+}}^{NH}}}$ essentially
the same as the vacuum probabilities. If the hierarchy is
inverted, then anti-neutrino probabilities undergo large changes
and neutrino probabilities remain the same. Thus ${\mathrm{\Delta
N_{\mu^-}}}$ and  ${\mathrm{\Delta N_{\mu^+}}}$ have opposite
signs leading to cancellations.  But ${\mathrm{\Delta N_{\mu^-}}}$
is larger by a factor $2.5-3$ because the neutrino-nucleon cross
sections are higher by this factor and hence the cancellation is
only partial. A similar reasoning holds for electron events. This
leads to hierarchy sensitivity in water \chr detectors. It is not
as good as that of detectors with charge discrimination capability
but the proposed megaton mass overrides this disadvantage and
provides the statistics necessary for a determination of the
hierarchy.

Since no simulation studies for atmospheric neutrinos are
available for \HK we assume the same detector characteristics as
in the \SK detector but with  increased statistics. Our results
are for an exposure of 1.8 \mtyrd, which corresponds to 3.3 years
of running time. In our calculation, we have put in a lepton
energy threshold of 1 GeV, since $\da$ driven matter effects arise
in this energy range and the determination of $\signda$ is better
achieved with higher energy neutrinos. This leads to a threshold
correction in the cross section~\cite{Petcov:2005rv}, which we
incorporate into our calculations. We use the detection
efficiencies of the \SK experiment~\cite{kajita,Ashie:2004mr} for
multi-GeV one-ring muon (both fully and partially contained
$\mu$-like) events and electron (e-like) events. The L/E analysis
of \SK muon-data has demonstrated the feasibility of
reconstructing
 the neutrino energy in a water \chr detector
from Monte Carlo simulation
by fitting it to the total
energy of the charged particles~\cite{Ashie:2004mr}.  Similarly the direction of the neutrino can
also be determined  from the reconstructed direction of the
muon~\cite{Ishitsuka:2004gd}. For fully contained (FC) (partially contained (PC)) multi-GeV muon events the
energy smearing is 17\% (24\%)  while the angular smearing is
$17^\circ (10^\circ)$. In our calculations for purposes of comparison with
magnetized iron calorimeter detectors, we use an overall 15\% energy smearing
and $10^\circ$ angular smearing for both types of events.
While this is somewhat optimistic, we will discuss in detail the extent to which the
sensitivity to hierarchy depends on smearing.

Both muon and electron events in this detector  have contributions
coming from background processes. The backgrounds in the FC events
are  due to cosmic ray muons, PMT flashes and neutron events from
the rock surrounding the detector~\cite{Ashie:2005ik}. For PC
events the  cosmic ray muons constitute the main background. These
backgrounds can, however, largely be eliminated during the  data
reduction procedure~\cite{Ashie:2005ik}. The remaining backgrounds
occur  due to (a) neutral current (NC) events and (b) $\nu_\mu$
(${\mathrm{\nu_e}}$) induced CC events for electron (muon) data
sample. The expected backgrounds in the case of atmospheric
neutrino interactions are estimated by the \SK Monte Carlo and are
given in~\cite{Ashie:2005ik}. For multi-GeV one-ring muon events
the contamination due to above processes is estimated to be about
$0.3\%$ whereas for multi-GeV one-ring electron events have a
background of about $10\%$ from NC events and about $7\%$ from
$\nu_\mu$ induced CC events.

%
%
\subsection{Magnetized Iron Detectors}

We consider a prototype magnetized iron detector along the lines of the India-based Neutrino
Observatory~\cite{Athar:2006yb} (\INOd).
This detector is expected to have a
modular structure with an initial running mass of 50 kT, building up to a
final mass of 100 kT.
It will consist of 140 layers of
iron plates about 6 cm thick, with gaps of
about 2.5 cm between them housing the active elements. These have been chosen to be RPC's (Resistive Plate Chambers),
made of glass or bakelite and containing a mixture of gases.
The (50 kT) structure is divided into three modules, with an overall
lateral size of 48 $\times$ 16 meters
and a height of 12 meters.
A magnetic field of about 1.3 Tesla will provide charge
discrimination capability to the detector.
We consider an exposure of 1.0 \mtyr, which corresponds to
10 years of running time.

The high density of iron renders this detector insensitive to
sub-GeV muons and electrons of any energy. In our calculations, we
have assumed a detection threshold of 1 GeV for muons and included
the corresponding threshold correction in the cross section. An
overall detection efficiency of 87\% and a charge identification
efficiency of 100\% is assumed~\footnote{This charge
identification efficiency is valid for neutrino energy $<<$ 1
TeV.}. The magnetic field allows charge identification  and thus
the interactions of $\nu_\mu$ and $\bar{\nu}_{\mu}$ can be studied
separately. Thus ${\mathrm{\Delta N_{\mu^-}}}$ and
${\mathrm{\Delta N_{\mu^+}}}$  defined in \eq~\ref{eq:Deltan} can
each be determined. The hierarchy sensitivity in this case depends
on the sum of the magnitudes of these quantities.  Therefore the
partial cancellation occuring for water \chr detectors does not
take place for charge discriminating detectors giving them  an
advantage over the former.

The energy and the direction of the neutrino in \INO can
be reconstructed from the muon track~\cite{Athar:2006yb}.
The energy of the neutrino is the sum total of the energy of the
muon and the hadrons. The latter is difficult to reconstruct
for individual hadrons. However, one may  use the hit multiplicity of charged particles distinct from the muon track
 to calibrate the
total energy of the hadrons in the event. It is reasonable to
assume energy smearing of 15\% and angular smearing of $10^\circ$
for this detector, which are the values we adopt in our numerical
work. As in the case of the water \chr detector, we study the
sensitivity of this detector's capability
 for hierarchy determination
in case its resolutions are different from those assumed by us.

Finally, we comment on the background signal in this detector. A
preliminary study using a GEANT based simulation of cosmic ray
muon background in \INO shows that these are unlikely to mimic the
signal~\cite{Indumathi:2004kd}. Other backgrounds can originate
due to NC interactions, such as ${\mathrm{\nu_x + d\;(u)
\rightarrow \nu_x + d\;(u) \;(+\; q \;\bar{q})}}$, where the
quarks in the final state can produce mesons along with other
hadrons. The decay of these mesons  produces secondary muons which
can contaminate the signal. However, simulations have shown that
the 6 cm thickness of the iron plates  is sufficient to absorb any
pions and kaons in the 1$-$10 GeV range before they  can decay. In
addition,  the oscillated $\nu_\tau$s can produce $\tau$ which can
decay to muons with a branching ratio of 17.36\% via $\nu_\tau
\rightarrow \mu^{-} + \bar{\nu_\mu} + \nu_\tau$ . However the
number of these secondary muons are expected to be small because
of the higher $\tau$ production threshold of 3.5 GeV.  Also these
muons are softer in energy and hence can be eliminated by suitable
energy cuts~\cite{Agarwalla:2005we}.

In \Tab~{\ref{table2}}, the comparative characteristics of the two detectors
\HK and \INO are listed.

%
%
\begin{table}
\begin{center}
\begin{tabular}
{ l | c | c }
\hline \hline
&&
\\
{\sf{Property}} & {{\HKd}} & {{\INOd}}
        \\
        &&
        \\
        \hline
     Detector Technology & Water \chr & Iron calorimeter \\
     Fiducial Volume     &  545 Kt  & 100 Kt   \\
     Exposure Time    & 3.3 yr & 10 yr  \\
     Energy Threshold    &  1 GeV & 1 GeV  \\
     Energy smearing    &  15\% &  15\% \\
     Angular Smearing & 10$^\circ$ & 10$^\circ$ \\
     Detection Efficiency & E-dependent & 87\% \\
    Charge Discrimination & No & Yes \\
     Muon Events & Yes & Yes \\
     Electron Events & Yes & No
          \\
\hline
\hline
\end{tabular}
\mycaption{Properties of the two detectors considered in our analysis, \HK and \INOd.}
\label{table2}
\end{center}
\end{table}

\section{Numerical Procedure}

The total number of CC events is obtained by  folding the relevant
incident neutrino fluxes with the appropriate disappearance and
appearance probabilities, relevant CC cross sections, the
efficiency for muon detection, the detector resolution, mass and
the  exposure time. The total CC cross section  used here is the
sum of quasi-elastic, single meson production and deep inelastic
cross sections. The cross sections for the water \chr detector are
taken from~\cite{Ashie:2005ik} and for the magnetized iron
detector are taken
from~\cite{Huber:2004ka,SK1,Paschos:2001np}. For the
incident atmospheric neutrino fluxes we use the tables
from~\cite{Honda:2004yz} where a 3-dimensional model is employed
for flux calculation~\footnote{The \INO facility is expected to be
housed at Pushep (lat: North $11.5^\circ$, long: East
$76.6^\circ$).
 The Honda fluxes in~\cite{Honda:2004yz} are calculated for \SK (lat: North $36.4^\circ$,
long: East $137.3^\circ$).
However, because of unavailability of the fluxes at the specific
\INO latitude we use the Honda fluxes given at the \SK site for
\INO.}.

For our analysis, we look at the neutrino energy range of 2$-$10
GeV and the cosine of the zenith angle ($\theta$) range of  -1.0
to -0.1. These  are divided into bins, and the $\mu^-$ event rate
in a specific energy bin with width $\mathrm{dE}$ and the solid
angle bin with width ${\mathrm{d \Omega}}$ is expressed as :
\begin{equation}
\rm{ \frac{d^2 N_{\mu}}{d \Omega \;dE} = \frac{1}{2\pi} ~\left[
\left(\frac{d^2 \Phi_\mu}{d \cos \theta \; dE}\right) P_{\mu\mu} +
\left(\frac{d^2 \Phi_e}{d \cos \theta \; dE}\right)
P_{e\mu}\right] ~\sigma_{CC} ~D_{eff} } \label{muevent}
\end{equation}
Here ${\mathrm{\Phi_{\mu,e}}}$ are the atmospheric fluxes
($\nu_\mu$ and ${\mathrm{\nu_e}}$), $\rm{\sigma_{CC}}$ is the
total CC cross section and $\rm{D_{eff}}$ is the detector
efficiency. The $\mu^+$ event rate is similar to the above
expression with the fluxes, probabilities and cross sections
replaced by those for anti-muons. Similarly, the ${\mathrm{e^-}}$
event rate in a specific energy and zenith angle bin is expressed
as follows:
\begin{equation}
\mathrm{ \frac{d^2 N_e}{d \Omega \;dE} = \frac{1}{2\pi}
~\left[\left( \frac{d^2 \Phi_{\mu}}{d \cos \theta \; dE}\right)
P_{\mu e} + \left(\frac{d^2 \Phi_e}{d \cos \theta \; dE}\right)
P_{e e} \right] ~\sigma_{CC} ~D_{eff} } \label{eevent}
\end{equation}
with the ${\mathrm{e^+}}$ event rate being expressed in terms of
anti-neutrino fluxes, probabilities and cross sections.

For the \HK analysis, the sum of ${\mathrm{\mu^- (e^-)}}$ events
and ${\mathrm{\mu^+ (e^+)}}$ events is taken to compute the total
muon (electron) event rate, since the detector is insensitive to
lepton charge.
 For the \INO analysis, however, the $\mu^-$ and $\mu^+$ event rates
are separately used given its charge identification capability.

%
\subsection{\label{sec:smearing} Energy and angular smearing:}

 We take into account the smearing in both energy and
zenith angle, assuming a Gaussian form of resolution
function, R. For energy, we use
\begin{equation}
\mathrm{ R(E_t,E_m) = \frac{1}{\sqrt{2\pi}\sigma} ~
\exp\left[-~\frac{(E_m - E_t)^2}{2 \sigma^2}\right]}
\label{esmear}
\end{equation}
Here, $\rm{E_m}$ and $\rm{E_t}$ denote the measured and true values
of energy respectively. The smearing width $\sigma$ is a fraction
of $\rm{E_t}$. Most of our calculations are performed assuming
this fraction to be $15\%$. We also calculate how our results vary
if this resolution fraction is reduced to $10\%$ or $5\%$.

The smearing function for the zenith angle is a bit more
complicated because the direction of incident neutrino is
specified by two variables: the polar angle ${\rm{\theta_t}}$ and
the azimuthal angle ${\rm{\phi_t}}$. We denote both these angles
together by ${\rm{\Omega_t}}$.
The measured direction of
the neutrino, with polar angle ${\rm{\theta_m}}$ and azimuthal
angle ${\rm{\phi_m}}$, which together we denote by ${\rm{\Omega_m}}$,
is expected to be within a cone of half angle $\Delta
\theta$ of the true direction.
Since the angular smearing is to be done in a cone
around the direction specified by ${\rm{(\theta_t,\phi_t)}}$,
we cannot assume the smearing function to be a function of
the polar angle only.
If we consider a small cone whose
axis is given by the direction ${\rm{\theta_t, \phi_t}}$, then
the set of directions within the cone have different polar
angles and azimuthal angles. Therefore, we need to construct
a smearing function which takes into account the change in
the azimuthal coordinates as well.
Such an angular smearing function is given by
\begin{equation}
\rm{ R(\Omega_t,\Omega_m) = N ~\exp \left[ -~ \frac{(\theta_t -
\theta_m)^2 + \sin^2 \theta_t ~(\phi_t - \phi_m)^2}{2 ~(\Delta
\theta)^2} \right] } \label{smearfac}
\end{equation}
Details of the computation of the normalization constant N
appearing in the above equation  are given in the
Appendix~\ref{app}.

The event rate with the smearing factors taken into account is
given by
\begin{equation}
\rm{ \frac{d^2 N_{\mu}}{d \Omega_m ~dE_m} = \frac{1}{2\pi}
~\int_{1}^{100} dE_t \int d\Omega_t~ R(E_t,E_m)~
R(\Omega_t,\Omega_m)~\left[\Phi_{\mu}^d \; P_{\mu\mu} +
\Phi_{e}^{d} \;P_{e\mu} \right]~ \sigma_{CC}~ D_{eff} }
\label{mueventsm}
\end{equation}
where we have  denoted $\rm{(d^2 \Phi/d \cos \theta \;dE)_\mu
\equiv \Phi_\mu^d}$ \etc. Strictly speaking, the range of
integration for the true energy of the  neutrino $\rm{E_t}$ should
be from $0$ to $\infty$. However, given the fact that we set a
lower threshold of 1 GeV for the lepton energy, our choice of the
lower
 limit of integration is dictated by the requirement that the neutrino be more energetic than the lepton.
Taking an upper limit of 100 GeV for the true energy is justified because
the probability of spillover of events from bins above 100 GeV to bins below 10 GeV due to smearing is insignificant, and the number
of events above 100 GeV is quite  small due to the steeply falling  neutrino flux.

%
%
\begin{figure*}[t]
\centerline{\includegraphics[width=4.5in]{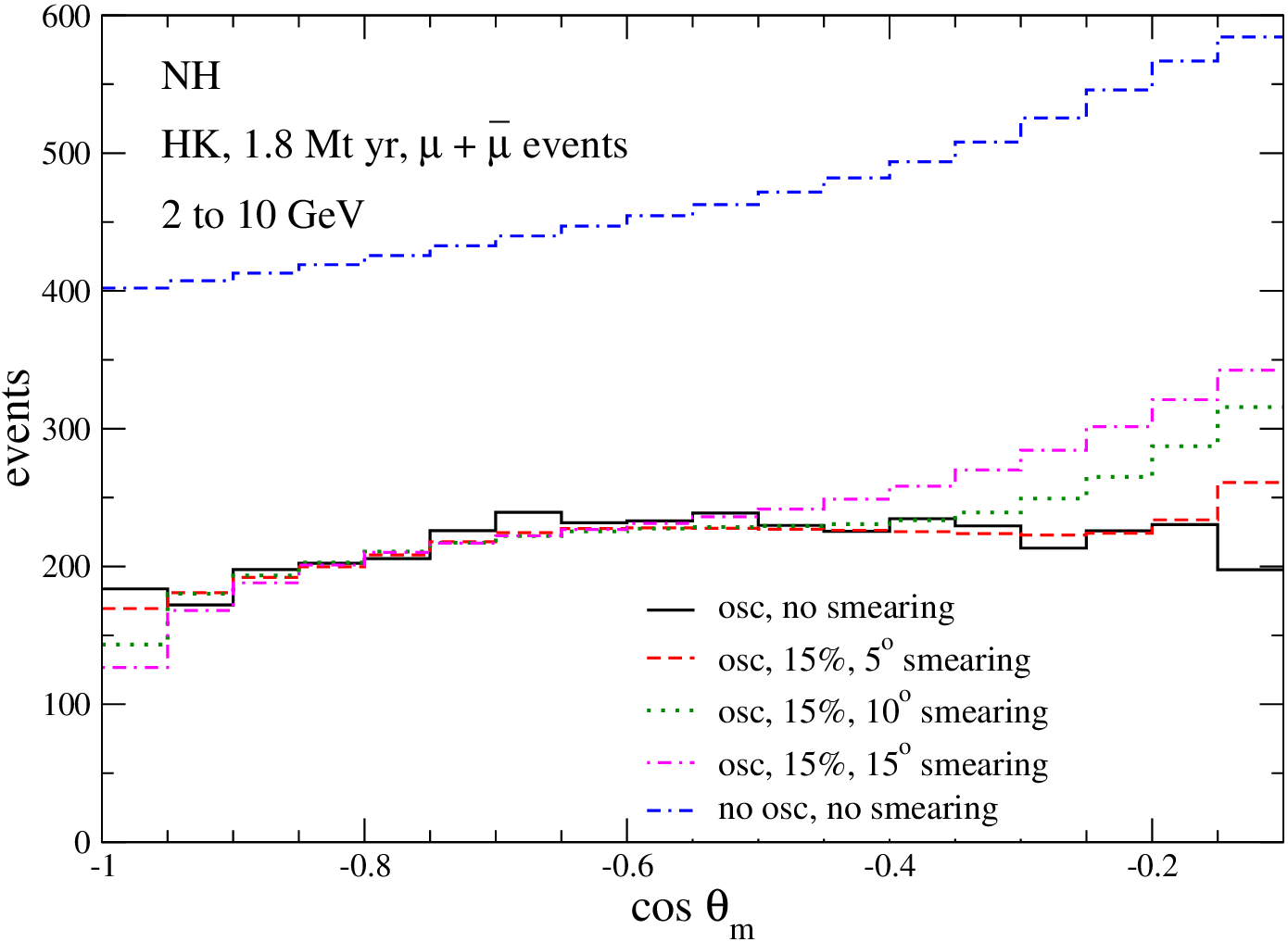}}
 \mycaption{The muon event rate for \HK (1.8 \mtyrd) with different values of angular
smearing and 15\% energy smearing, as well as without smearing
with and without oscillations, plotted versus ${\mathrm{\cos
\theta_m}}$ taking 18 bins in the range -1.0 to -0.1. The energy
range taken is 2 to 10 GeV, and the hierarchy is taken to be
normal. }
    \label{fig1e}
\end{figure*}

\begin{figure}[h]
\centerline
{
\epsfxsize=7cm\epsfysize=13.50cm
                     \epsfbox{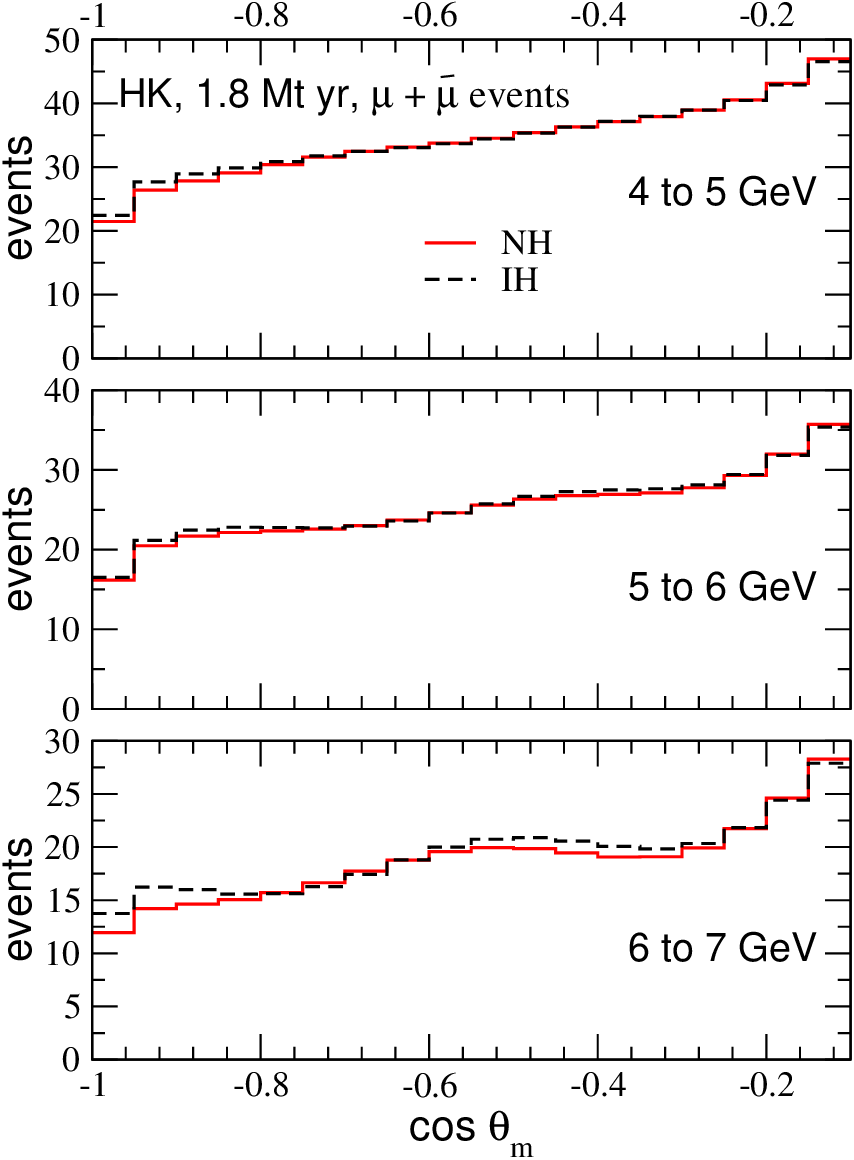}
\hspace*{15.0ex}
\epsfxsize=7cm\epsfysize=13.50cm
                     \epsfbox{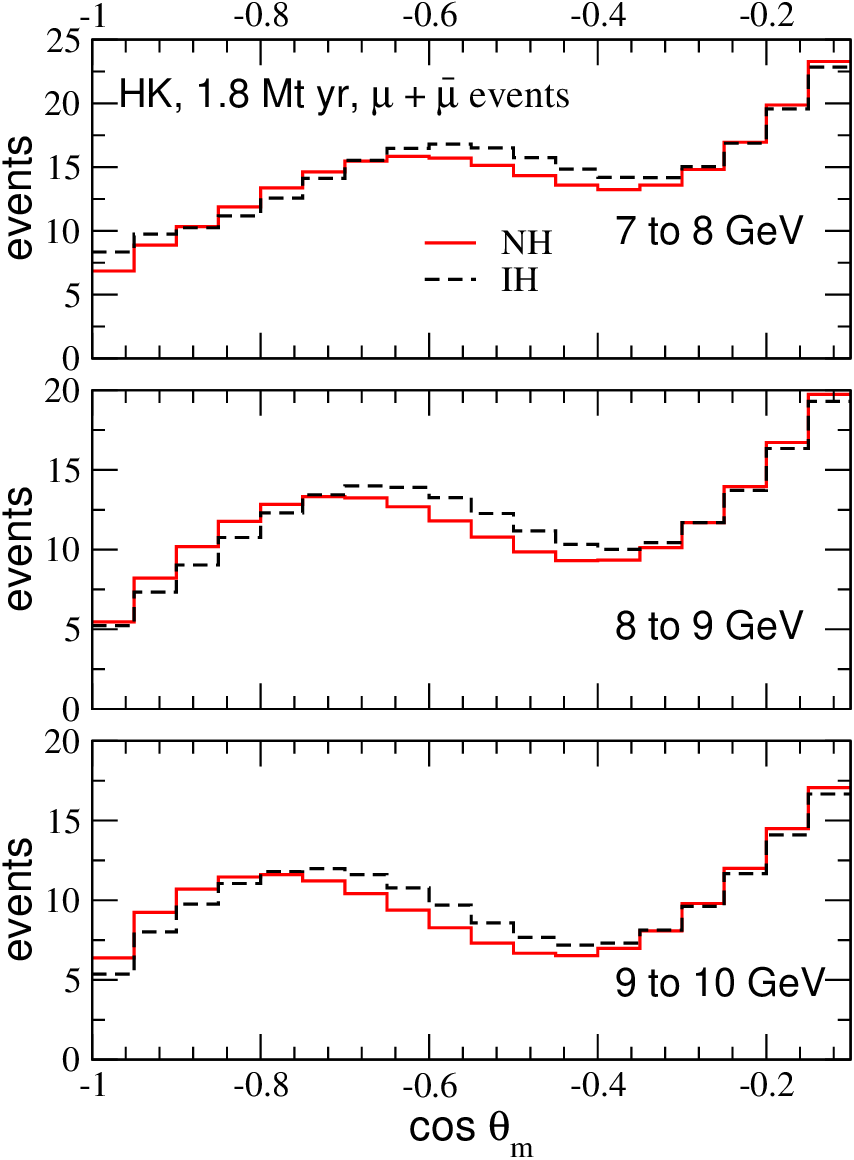}
                     }
\mycaption{ The muon event rate for \HK (1.8 \mtyrd) with
10$^\circ$ angular smearing and 15\% energy smearing plotted
versus ${\mathrm {\cos \theta_m}}$ taking 18 bins in the range
-1.0 to -0.1, for 6 different energy bins in the range 4 to 10
GeV, for both normal and inverted hierarchy. } \label{fig1f}

\end{figure}

In \fig~{\ref{fig1e}}, we show the effect of angular smearing on
distribution of muon events in \HK assuming a NH. The figure also
shows the ${\mathrm{\cos \theta_m}}$ distribution for the two
special cases of {\it no} smearing {\it without} and {\it with}
oscillations. It demonstrates the washing out of oscillatory
behaviour in the event distribution as the angular smearing width
is increased. Note that the distribution with oscillation and
without smearing shows some distortion in its shape compared to
the distribution without oscillation. When the angular smearing is
introduced, this oscillatory  distortion is washed out
progressively as the value of smearing is increased.
 For smearing with a large angular resolution of 15$^\circ$, the distribution
with oscillations resembles the unoscillated distribution in its shape.

\fig~{\ref{fig1f}} shows the muon event distribution in \HK for NH
and IH for six energy bins of width 1 GeV each in the energy range
of 4$-$10 GeV. From these figures it is clear that below the
resonant energy, the difference between the normal and inverted
hierarchy distributions is negligible (note that
${\mathrm{E_{res}}}
> 6$ GeV for all baselines considered here, as described in
\Tab~{\ref{eresvalues}}). Around resonance, \ie~from the 6$-$7 GeV
energy bin, the difference begins to be significant. A little
above the resonance, the event numbers fall off, but a reasonable
difference between NH and IH distributions persists, which
contributes significantly to the hierarchy sensitivity.

\vskip.5cm
\noindent
{\bf{$\chi^2$ analysis}}
\vskip.5cm

We study the $\chi^2$ sensitivity to the mass hierarchy for
various different values of energy and angular resolution. For
\HKd, we use a binned distribution of muon as well as electron
events in various ${\mathrm{E_m}}$ and ${\mathrm{\cos \theta_m}}$
bins. For \INOd, we use similar binning for $\mu^-$ and $\mu^+$
events.

From \fig~{\ref{fig1f}} we note that each bin contains $\geq 5$
events. Hence Gaussian error analysis may be used.
In the limit when only statistical errors are taken into account,
the standard Gaussian definition of binned $\chi^2$ is:
\begin{equation}
{\mathrm{ \chi^2_{stat} = \sum_{i=E_m bins} \ \
 \sum_{j=\cos \theta_m
bins} ~\frac{ \left[~N_{ij}^{ex}-N_{ij}^{th}~\right]^2}
{N_{ij}^{ex}}
}} \label{chisqstat}
\end{equation}
Here, ${\mathrm N_{ij}^{ex}}$ is the experimental and ${\mathrm
N_{ij}^{th}}$ is theoretical number of events in the
${\mathrm{ij^{th}}}$ bin.

However, in addition to the statistical uncertainties, one also
needs to take into account various theoretical and systematic
uncertainties. In our analysis  we include the uncertainties
coming from
\begin{itemize}
\item A flux normalization error of 20$\%$,
\item A tilt
factor~\cite{Gonzalez-Garcia:2004wg} which takes into account the
deviation of the atmospheric fluxes from a power law,
\item A
zenith angle dependence uncertainty of 5$\%$,
\item An overall
cross section uncertainty of 10$\%$,
\item An overall systematic uncertainty of 5$\%$.
\end{itemize}
These uncertainties are included using the method of pulls
described
in~\cite{Fogli:2002pt,Fogli:2003th,Gonzalez-Garcia:2004wg}. This
method allows us to take into account the various uncertainties in
theoretical inputs and experimental systematics in a simple,
straight-forward way.

In this method, the uncertainty in fluxes and cross sections and
the systematic uncertainties are taken into account by allowing
these inputs to deviate from their standard values in the
computation of ${\mathrm{ N^{th}_{ij} }}$. Let the ${\mathrm{
k^{th} }}$ input deviate from its standard value by ${\mathrm{
\sigma_k \;\xi_k }}$, where ${\mathrm{ \sigma_k }}$ is its
uncertainty. Then the value of ${\mathrm{ N^{th}_{ij} }}$ with the
changed inputs is given by
\begin{equation}
{\mathrm{ N^{th}_{ij} =  N^{th}_{ij}(std) + \sum^{npull}_{k=1}\;
c_{ij}^k \;\xi_k }} \label{cij}
\end{equation}
where ${\mathrm{ N^{th}_{ij}(std) }}$ is the theoretical rate for
bin ${\mathrm{ij}}$, calculated with the standard values of the
inputs and npull is the number of sources of uncertainty, which in
our is case is 5.
The ${\mathrm{ \xi_k }}$'s are called the ``pull" variables and
they determine the number of ${\mathrm{ \sigma}}$'s by which the
${\mathrm{ k^{th} }}$ input deviates from its standard value. In
\eq~\ref{cij}, ${\mathrm{ c_{ij}^k }}$ is the change in ${\mathrm{
N^{th}_{ij} }}$ when the ${\mathrm{ k^{th} }}$ input is changed by
${\mathrm{ \sigma_k }}$ (\ie~by 1 standard deviation). The
uncertainties in the inputs are not very large. Therefore, in
\eq~\ref{cij} we only considered the changes in ${\mathrm{
N^{th}_{ij} }}$ which are linear in ${\mathrm{ \xi_k }}$. Thus we
have a modified $\chi^2$ defined by
\begin{equation}
\mathrm{ {\chi^2(\xi_k)} = \sum_{i,j}\;
\frac{\left[~N_{ij}^{th}(std) \;+\; \sum^{npull}_{k=1}\;
c_{ij}^k\; \xi_k - N_{ij}^{ex}~\right]^2}{N_{ij}^{ex}} +
\sum^{npull}_{k=1}\; \xi_k^2  } \label{chisqxik}
\end{equation}
where the additional term ${\rm{\xi_k^2}}$ is the penalty imposed
for moving ${\mathrm{k^{th}}}$ input away from its standard value
by ${\rm{\sigma_k \;\xi_k}}$. The $\chi^2$ with pulls, which
includes the effects of all theoretical and systematic
uncertainties, is obtained by minimizing ${\rm{\chi^2(\xi_k)}}$,
given in \eq~\ref{chisqxik}, with respect to all the pulls
${\rm{\xi_k}}$:
\begin{equation}
{\mathrm{ \chi^2_{pull} = Min_{\xi_k}~ \left[~
\chi^2(\xi_k)~\right] }} \label{chisqpull}
\end{equation}

%
%
\subsection{Optimization of number of bins}

We do binning in the measured energy $\rm{E_m}$ and the cosine of
the measured zenith angle (which is the measured polar angle of
the incident neutrino) ${\mathrm{\cos \theta_m}}$. For
optimization purposes, we compute the $\chi^2$ with pull for
various different choices of the number of  energy and zenith
angle bins. We found that the sensitivity for muon events improves
with an increase in the number of bins, and is optimal for about
15 ${\mathrm{\cos \theta_m}}$ bins and 15 energy bins. For the
electron events, the optimization occurs for a lower number of
bins. This difference can be understood from the behaviour of
$\pee$, $\pmue$ and $\pmumu$ in \figs~\ref{fig1} and \ref{fig2}.
In the resonance energy range of 4$-$6 GeV, oscillation
probabilities involving ${\mathrm{\nu_e}}$ are relatively less
sharply oscillating, whereas the $\pmumu$ oscillates rapidly. Thus
a finer binning in energy is needed for muon events to capture
these oscillations, whereas a much coarser energy binning is
enough to capture the change due to matter effect in electron
events. However, because the energy resolution for a water \chr
detector gets poorer at higher energies in the GeV range, a very
fine binning in energy for ${\mathrm{E_m > 2}}$ GeV is not
realistic. Hence we consider a bin division into 8
${\mathrm{E_m}}$ bins in the range 2$-$10 GeV and 18
${\mathrm{\cos \theta_m}}$ bins in the range -1.0 to -0.1. The
results given in the subsequent sections are with this binning.

%
\subsection{Marginalization over Neutrino Parameters}

In  order to determine the optimum number of bins we have used the
${\rm{\chi^2_{pull}}}$ from \eq~\ref{chisqpull} where
only the uncertainties in inputs such as fluxes and cross sections
are taken into account, holding the values of
the oscillation parameters fixed in the calculation of both
$\rm{N^{ex}_{ij}}$ and $\rm{N^{th}_{ij}}$.
However, in general,
the values of the mass-squared differences $|\da|$ and
$\ds$ and the mixing angles $\theta_{12}$, $\theta_{23}$ and
$\theta_{13}$
can vary over a range that reflects the uncertainty in our knowledge.
Holding their values as fixed in computing both ${\rm N^{ex}_{ij}}$ and
${\rm N^{th}_{ij}}$
is tantamount  to assuming that they  are known to
infinite precision, which is not  realistic.
To take into account the uncertainties in these parameters,
we define the {\bf{marginalized}} $\chi^2$ for hierarchy sensitivity as,
\begin{eqnarray}
{\mathrm{ \chi^2_{min}}}  &=& {\mathrm{ Min \left[~ \chi^2 (\xi_k)
~+~ \left(\; \frac{|\da|^{true} - |\da|}{\sigma\;(|\da|)}
\;\right)^2 \right.}}
\nonumber \\
&&{\mathrm{ \left. ~+~ \left(\; \frac{\sin^2 2\theta_{23}^{true} -
\sin^2 2\theta_{23}}{\sigma\;(\sin^2 2 \theta_{23})} \;\right)^2
~+~ \left(\; \frac{\sin^2 2\theta_{13}^{true} - \sin^2 2
\theta_{13}}{\sigma\;(\sin^2 2\theta_{13})}\;\right)^2 ~\right] }}
\label{chisqfinal}
\end{eqnarray}
${\rm{\chi^2(\xi_k)}}$ in the above equation, is computed according
to the definition given in \eq~\ref{chisqxik}.

We use the following procedure for our analysis.
\begin{itemize}

\item We simulate the number of events in 8 bins in the measured
energy ${\rm{E_m}}$ and  18 bins in the measured zenith angle
${\rm{\cos\theta_m}}$ for a set of ``true values" for the six
neutrino parameters -- $|\da|$, $\theta_{23}$, $\theta_{13}$,
$\ds$, $\theta_{12}$ and $\dcp$ and for a ``true hierarchy". This
is our ``experimental data" -- ${\rm N^{ex}_{ij}}$.

\item In the case of $\ds$, $\theta_{12}$, $|\da|$ and
$\theta_{23}$, the {\it true values} are the current best-fit
values. The true value of $\dcp$ is assumed to be zero in our
analysis~\footnote{The oscillation probabilities have only a weak
dependence on $\dcp$ for values of $\sin^2 2 \theta_{13}$ which
will be measurable in the forthcoming reactor neutrino experiments
$(\geq 0.05)$. In the next section, we will explicitly show that
for both water \chr and magnetized iron detectors, $\chi^2$ has
only a weak dependence on $\dcp$. Therefore we set $\dcp =0$.}.

\item Matter effects, which are crucial for making a distinction
between the hierarchies, are proportional to $\sin^2 2
\theta_{13}$. At present there is only an upper limit on $\sin^2
2\theta_{13}$ ($<0.15$). Therefore, we compute the $\chi^2$ for
various different ``true" input values of this parameter in its
permissible range.

\item In order to test at what statistical significance the
``wrong hierarchy" can be disfavoured, we calculate the
theoretical expectation in each bin -- ${\rm N^{th}_{ij}}$
assuming the ``wrong hierarchy".

\item During this calculation  of theoretical event rates we fix
the solar parameters $\ds$ and $\theta_{12}$ at their best-fit
values. Since the solar parameters have only marginal effect on
the probabilities for the energies and pathlengths relevant for
us, our results will not change significantly if these two
parameters are allowed to vary in their currently allowed range.
For the same reason we keep  the CP phase $\dcp$ fixed at its true
value which we have taken as zero in this calculation.

\item However,  we allow the parameters $\sin^2 2 \theta_{13}$,
$\sin^2 2 \theta_{23}$ and $|\da|$ to vary within the following
ranges :
\\
\begin{enumerate}
  \item $|\da|$ is allowed to vary
  in the range $2.35 \times 10^{-3} - 2.65
  \times 10^{-3}$ eV$^2$.
\item
$\sin^2 2 \theta_{23}$ is varied between 0.95 and 1.0. However, $\pmue$
and $\pmumu$ in matter are dependent on $\sin^2 \theta_{23}$.
For $\sin^2 2\theta_{23} < 1$ there exist two allowed values of
$\theta_{23}$ (the so called octant ambiguity). In our calculation we
consider both values. Hence we consider a range $0.4 < \sin^2\theta_{23} < 0.6$.
\item
$\sin^2 2 \theta_{13}({\rm true})$ is varied from 0.0 to 0.15.
The current 3$\sigma$ bound is $\sin^2 2 \theta_{13} < 0.15$~\cite{Schwetz:2006dh}.
\end{enumerate}

\item
In computing ${\rm{\chi^2_{min}}}$, we have
added the {\it priors} for the neutrino parameters
which puts a penalty for moving away from the true value.
Shifting further
from the {\it true value} of a parameter, would worsen the fit
of the experiment which measured that parameter. By adding the
{\it priors} we are effectively minimizing $\chi^2$ of {\it our data}
together with those of the experiments measuring the
neutrino parameters.
The results, obviously, depend on the choice of true parameter values.

In the expression for ${\rm{\chi^2_{min}}}$, the prior for the mixing
angle $\theta_{23}$ is given in terms of $\sin^2 2 \theta_{23}$.
This is valid because the quantity which will be measured in
future $\nu_\mu$ disappearance
experiments is $\sin^2 2 \theta_{23}$ and the priors are added
to take into account the fit to the data which made the measurements.

In \eq~\ref{chisqfinal},  $\sigma$ denotes 1$\sigma$ errors. We
use $2 \%$ error for $|\da|$ and $\sin^2 2\theta_{23}$, which can
be achieved in future long baseline
experiments~\cite{Itow:2001ee}. For $\sigma(\sin^2 2\theta_{13})$
we use 0.02~\cite{Huber:2004ug}.
\end{itemize}

Our $\chi^2$ is thus marginalized  over the three parameters
$\sin^2 2 \theta_{13}$, $\sin^2 2 \theta_{23}$ and $|\da|$ in
order to determine ${\rm{\chi^2_{min}}}$ which shows how different
are the predictions of the the ``wrong hierarchy" from those of
the ``true hierarchy". ``Wrong hierarchy" is then taken to be
ruled out at ${\rm{p\sigma}}$ if ${\rm{\chi^2_{min}}} \geq
{\rm{p^2}}$ for all allowed values of $\theta_{13}$, $\theta_{23}$
and $|\da|$.

%
%
In our calculations, we took the density profile of the earth to
be the PREM profile. There are, of course, some uncertainties in
the values of the densities given in this profile. We checked that
a $10\%$ change in the density leads to a negligible change (less
than $5\%$ change) in the minimum $\chi^2$. Therefore, we have not
taken the uncertainties in the density profile into account
explicitly.

\begin{figure*}[t]
   \centerline{\includegraphics[width=6.5in]{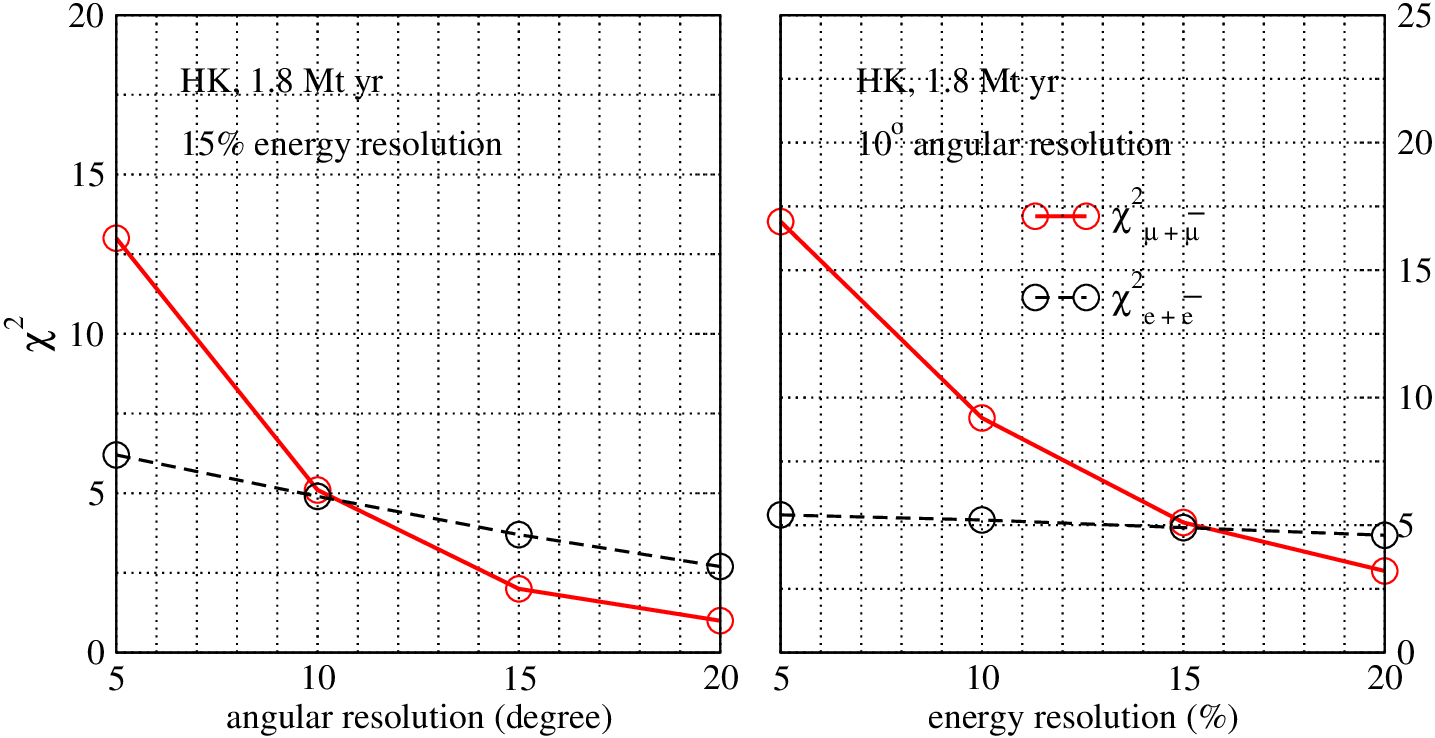}}
    \mycaption{Values of fixed parameter $\chi^2$ versus angular resolution
(left panel) and energy resolution (right panel) for \HK (1.8
\mtyrd). Plots are given separately for $\chi^2_{\mu+{\bar{\mu}}}$
and ${\mathrm{\chi^2_{e+{\bar{e}}}}}$. For the left panel, an
energy resolution of 15\% is assumed, whereas for the right panel,
an angular resolution of $10^\circ$ is assumed. No marginalization
over neutrino parameters is done. }
    \label{fig5}
\end{figure*}

\begin{figure*}[t]
   \centerline{\includegraphics[width=6.5in]{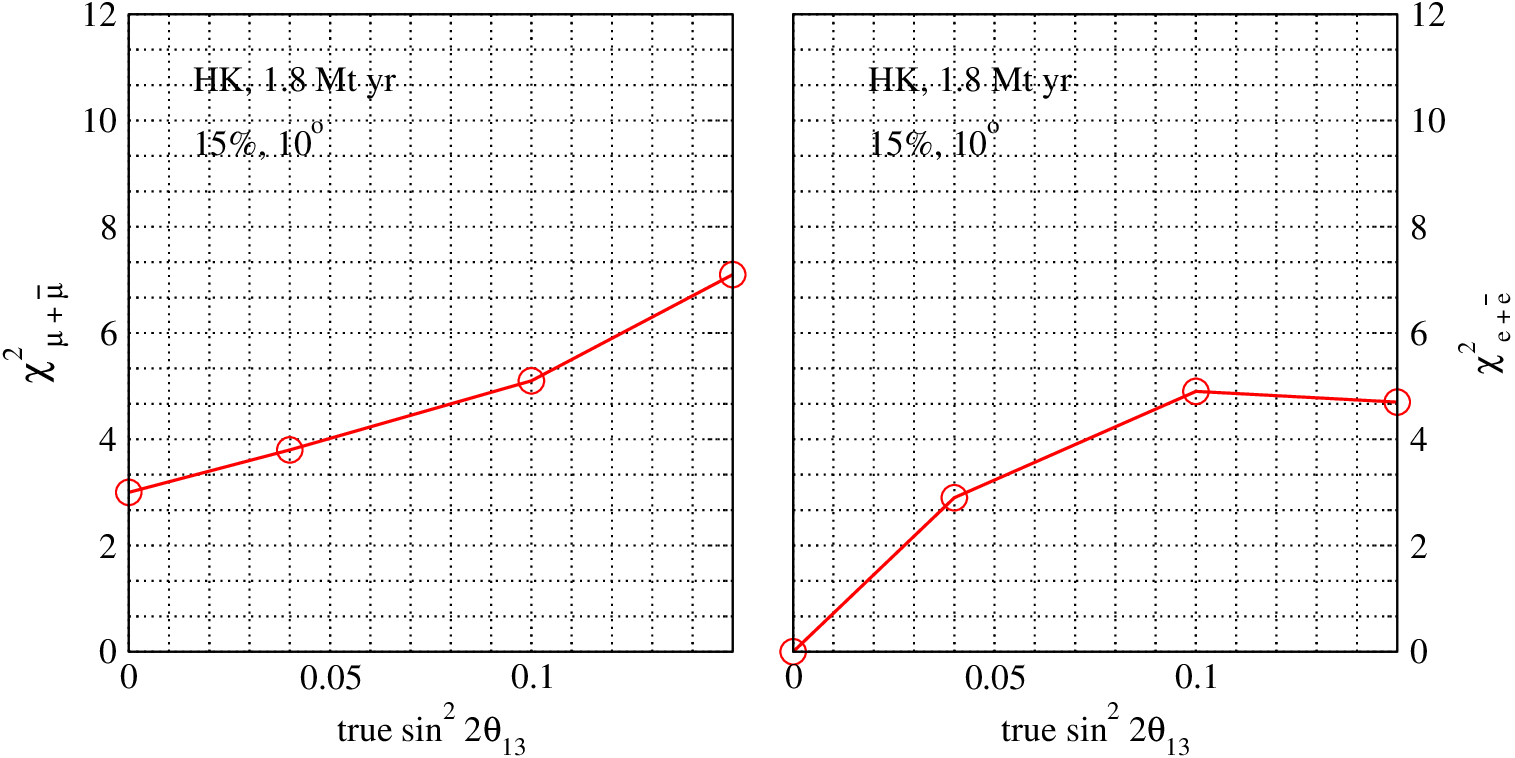}}
    \mycaption{Values of fixed parameter $\chi^2$ versus
the input (true) value of $\sin^2 2 \theta_{13}$ for \HK (1.8
\mtyrd), assuming {\bf NH} to be the true hierarchy. Left panel
shows $\chi^2_{\mu+{\bar{\mu}}}$ and right panel shows
${\mathrm{\chi^2_{e+{\bar{e}}}}}$. An energy resolution of $15 \%$
and an angular resolution of $10^\circ$ are assumed. No
marginalization over neutrino parameters is done.
%
}
    \label{fig6}
\end{figure*}


\begin{figure*}[t]
   \centerline{\includegraphics[width=6.5in]{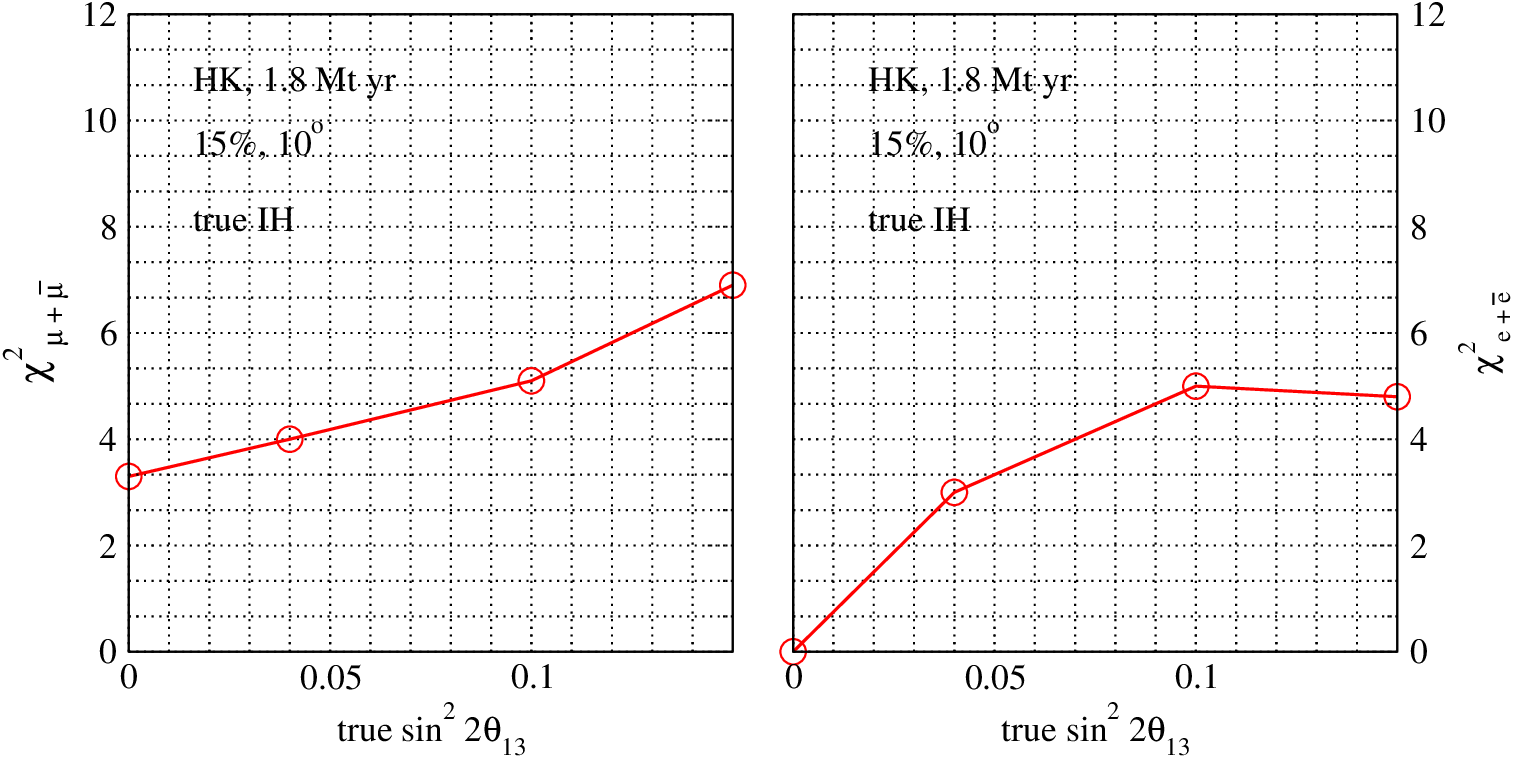}}
    \mycaption{Values of fixed parameter $\chi^2$ versus
the input (true) value of $\sin^2 2 \theta_{13}$ for \HK (1.8
\mtyrd), assuming {\bf IH} to be the true hierarchy.
Left panel shows $\chi^2_{\mu+{\bar{\mu}}}$ and right panel shows
${\mathrm{\chi^2_{e+{\bar{e}}}}}$.
An energy resolution of $15\%$ and an angular resolution of
$10^\circ$ are assumed. No marginalization over neutrino
parameters is done.
%
}
    \label{fig7}
\end{figure*}


\begin{figure*}[t]
   \centerline{\includegraphics[width=6.5in]{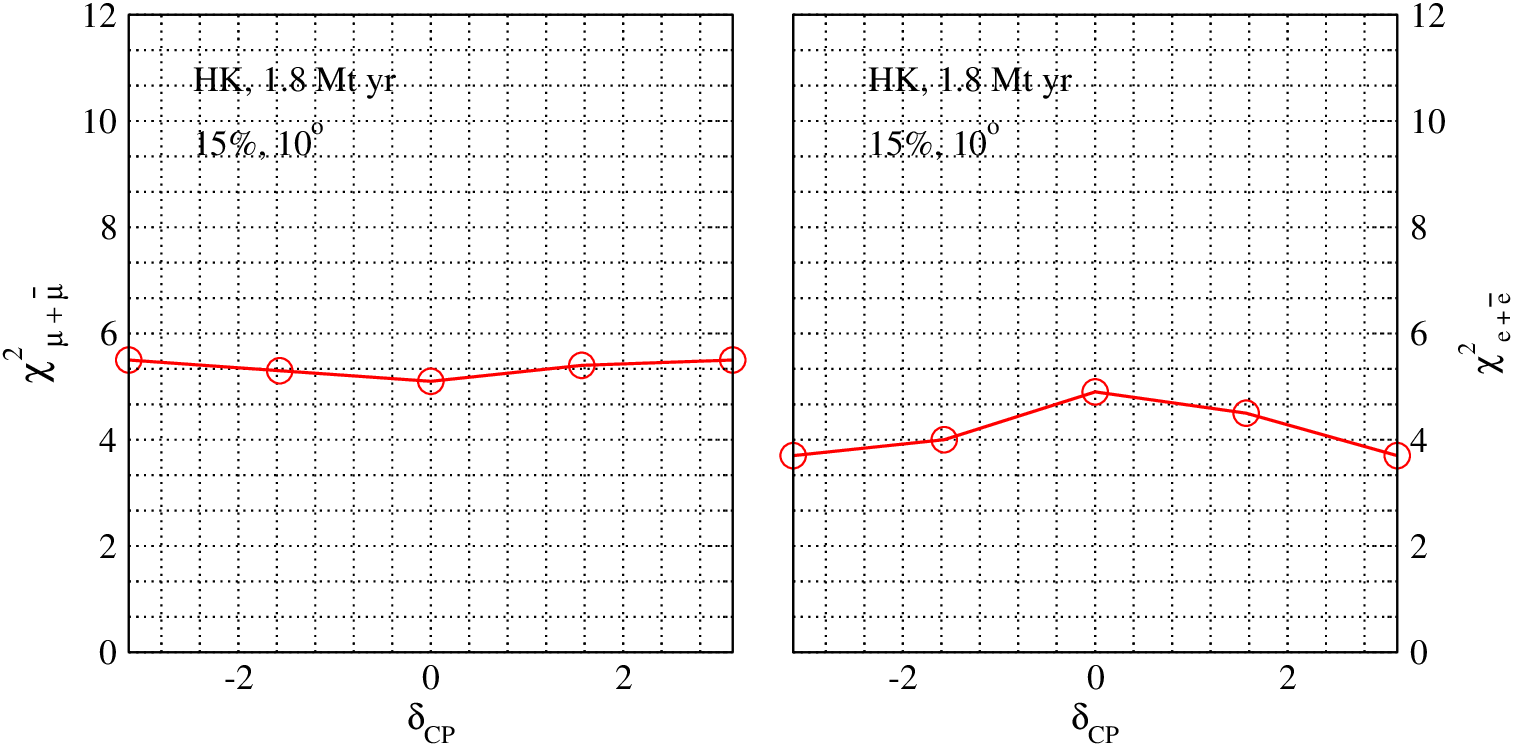}}
   %
    \mycaption{Values of fixed parameter $\chi^2$ versus
the input (true) value of $\dcp$ for \HK (1.8 \mtyrd).
Left panel shows $\chi^2_{\mu+{\bar{\mu}}}$ and right panel shows
${\mathrm{\chi^2_{e+{\bar{e}}}}}$.
An energy resolution of $15\%$ and an angular resolution of
$10^\circ$ are assumed. {\bf NH} is assumed to be true hierarchy.
No marginalization over neutrino parameters is done.
}
    \label{fig8}
\end{figure*}
\begin{figure*}[t]
   \centerline{\includegraphics[width=4.5in]{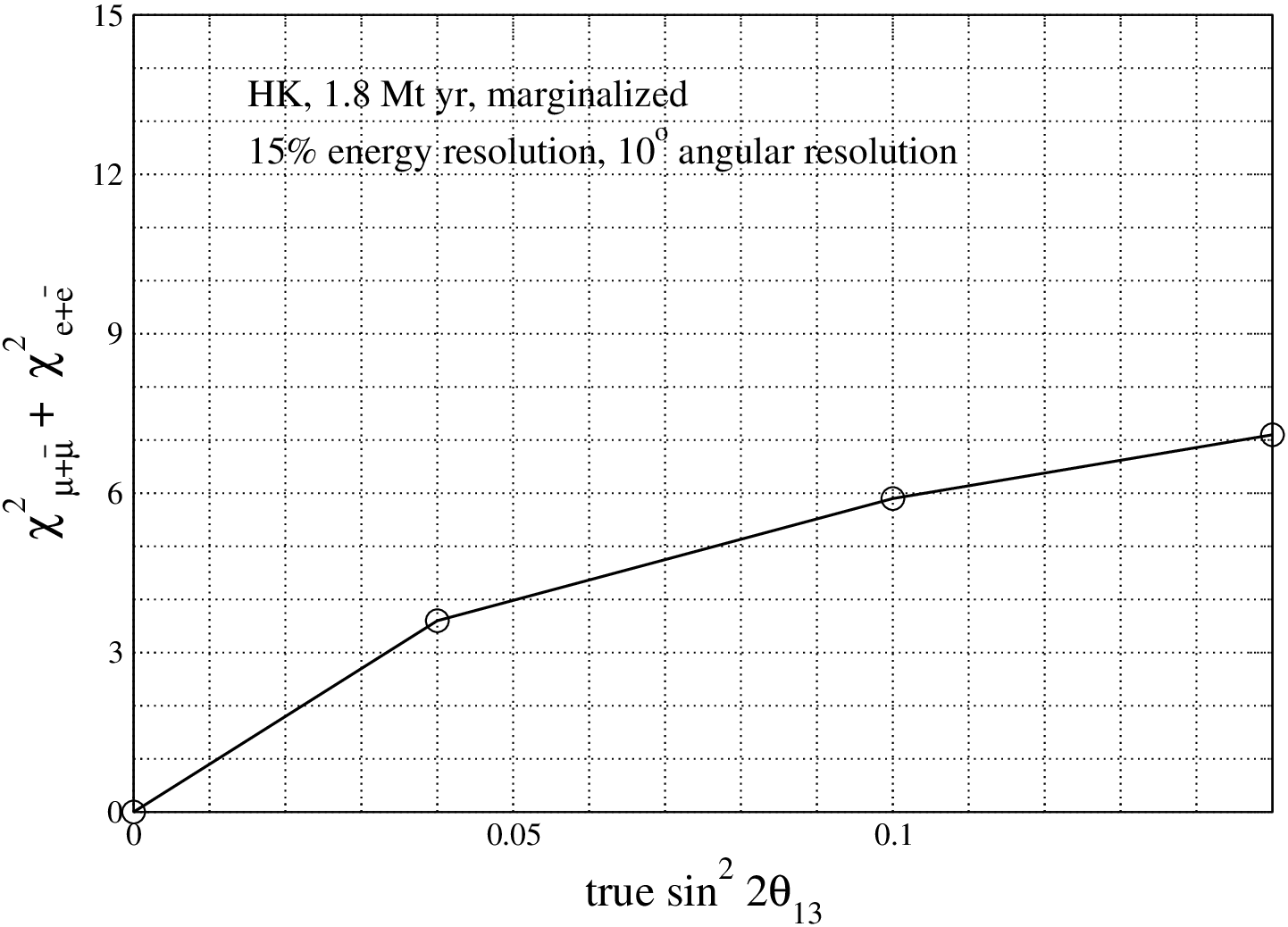}}
    \mycaption{Values of marginalized $\chi^2$ versus
the input (true) value of $\sin^2 2 \theta_{13}$ for \HK (1.8
\mtyrd). Shown is the sum of $\chi^2_{\mu+{\bar{\mu}}}$ and
${\mathrm{\chi^2_{e+{\bar{e}}}}}$.
An energy resolution of 15\% and an angular resolution of
$10^\circ$ are assumed.
}
    \label{fig9}
\end{figure*}



\begin{figure*}[t]
   \centerline{\includegraphics[width=6in]{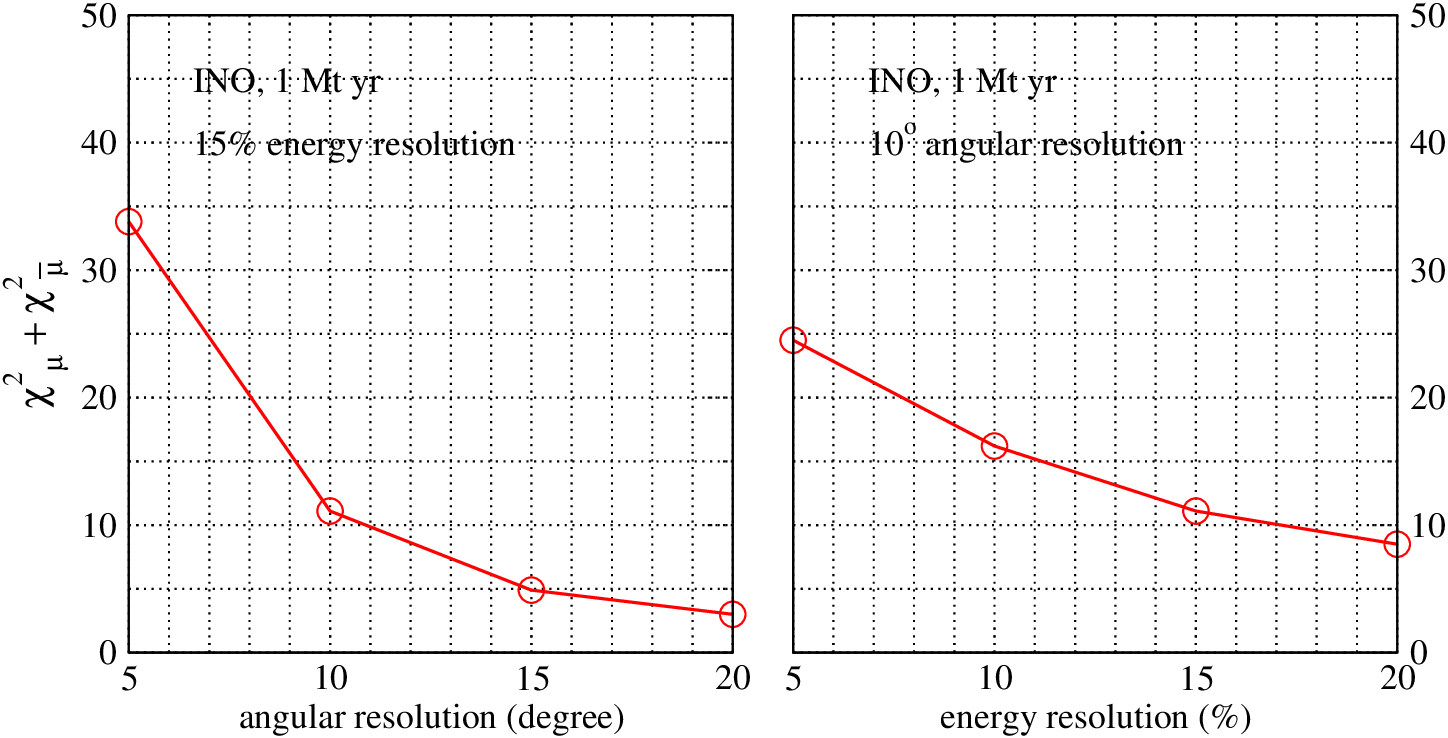}}
    \mycaption{Values of fixed parameter $\chi^2$ versus angular resolution
(left panel) and energy resolution (right panel) for \INO (1 \mtyrd).
Shown is the sum of $\chi^2_{\mu}$ and
$\chi^2_{\bar{\mu}}$.
For the left panel, an energy resolution of $15 \%$ is assumed,
whereas for the right panel, an angular resolution of $10^\circ$
is assumed. No marginalization over neutrino parameters is done.
}
    \label{fig10}
\end{figure*}



\begin{figure*}[t]
   \centerline{\includegraphics[width=6in]{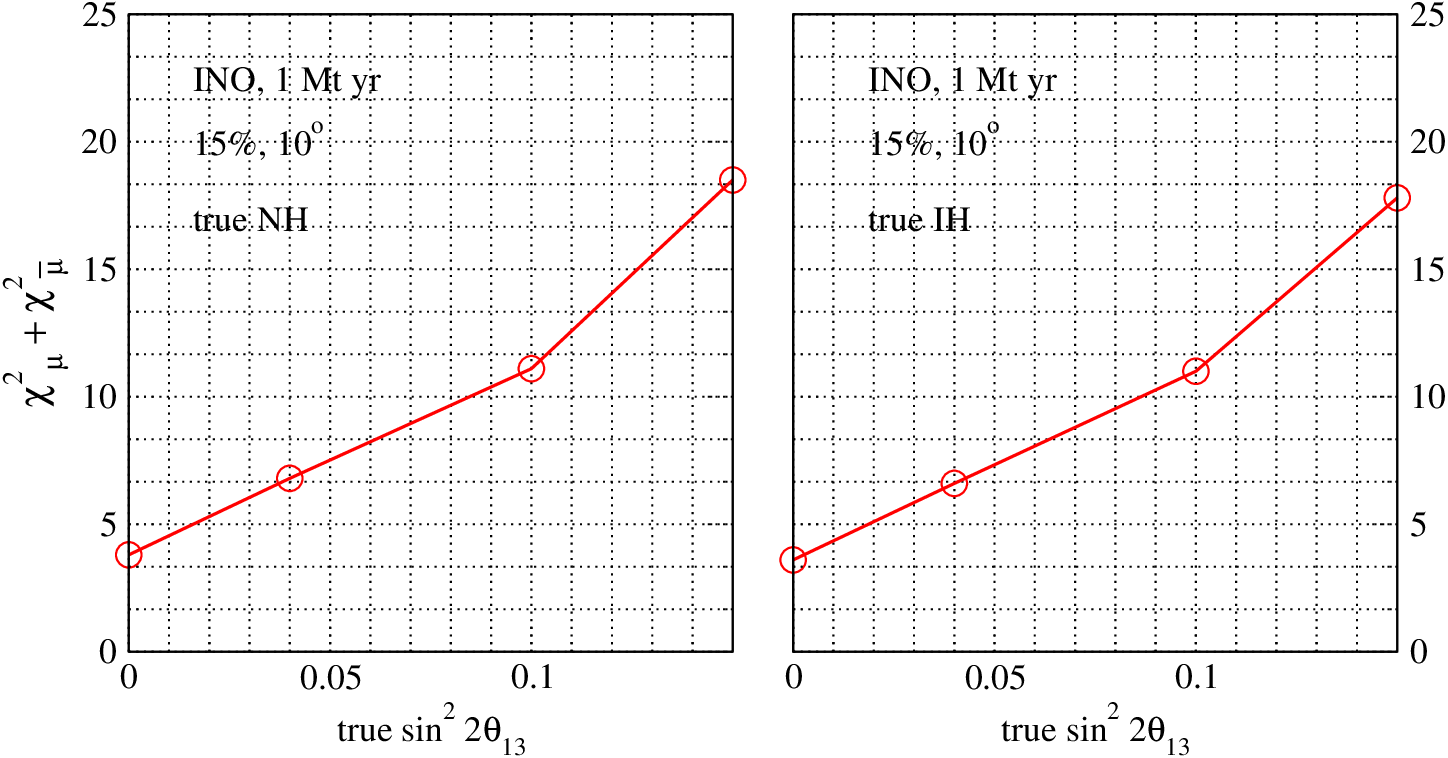}}
    \mycaption{Values of fixed parameter $\chi^2$ versus
the input (true) value of $\sin^2 2 \theta_{13}$ for \INO (1
\mtyrd), assuming {\bf NH} (left panel) and {\bf IH} (right panel)
to be the true hierarchy. Shown is the sum of $\chi^2_{\mu}$ and
$\chi^2_{\bar{\mu}}$.
An energy resolution of $15\%$ and an angular resolution of
$10^\circ$ are assumed. No marginalization over neutrino
parameters is done.
}
    \label{fig11}
\end{figure*}



\begin{figure*}[t]
   \centerline{\includegraphics[width=4.5in]{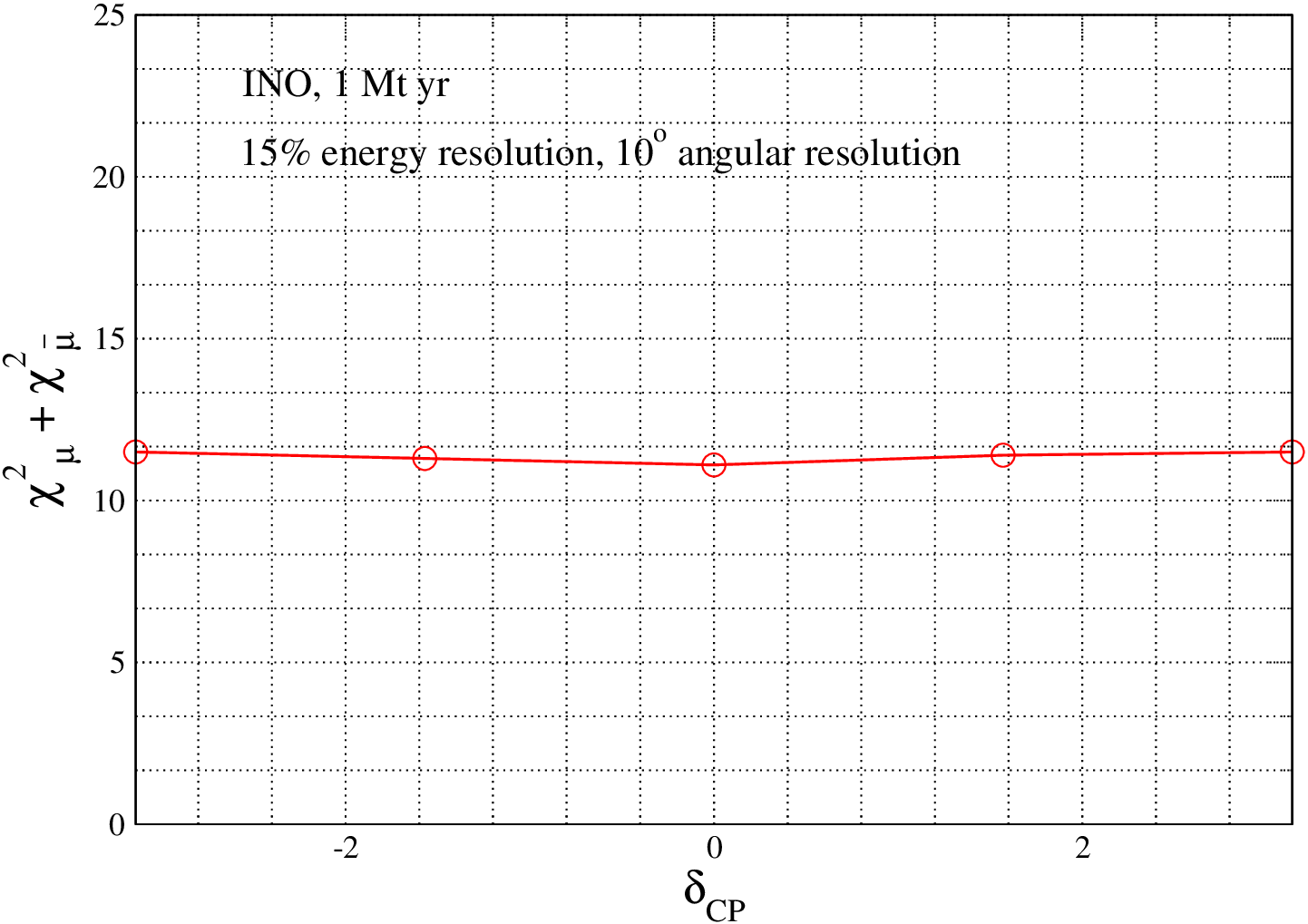}}
   %
    \mycaption{Values of fixed parameter $\chi^2$ versus
the input (true) value of $\dcp$ for \INO (1 \mtyrd). Shown is the
sum of $\chi^2_{\mu}$ and $\chi^2_{\bar{\mu}}$.
An energy resolution of $15\%$ and an angular resolution of
$10^\circ$ are assumed. {\bf NH} is assumed to be true hierarchy.
No marginalization over neutrino parameters is done.
}
    \label{fig13}
\end{figure*}



\begin{figure*}[t]
   \centerline{\includegraphics[width=4.5in]{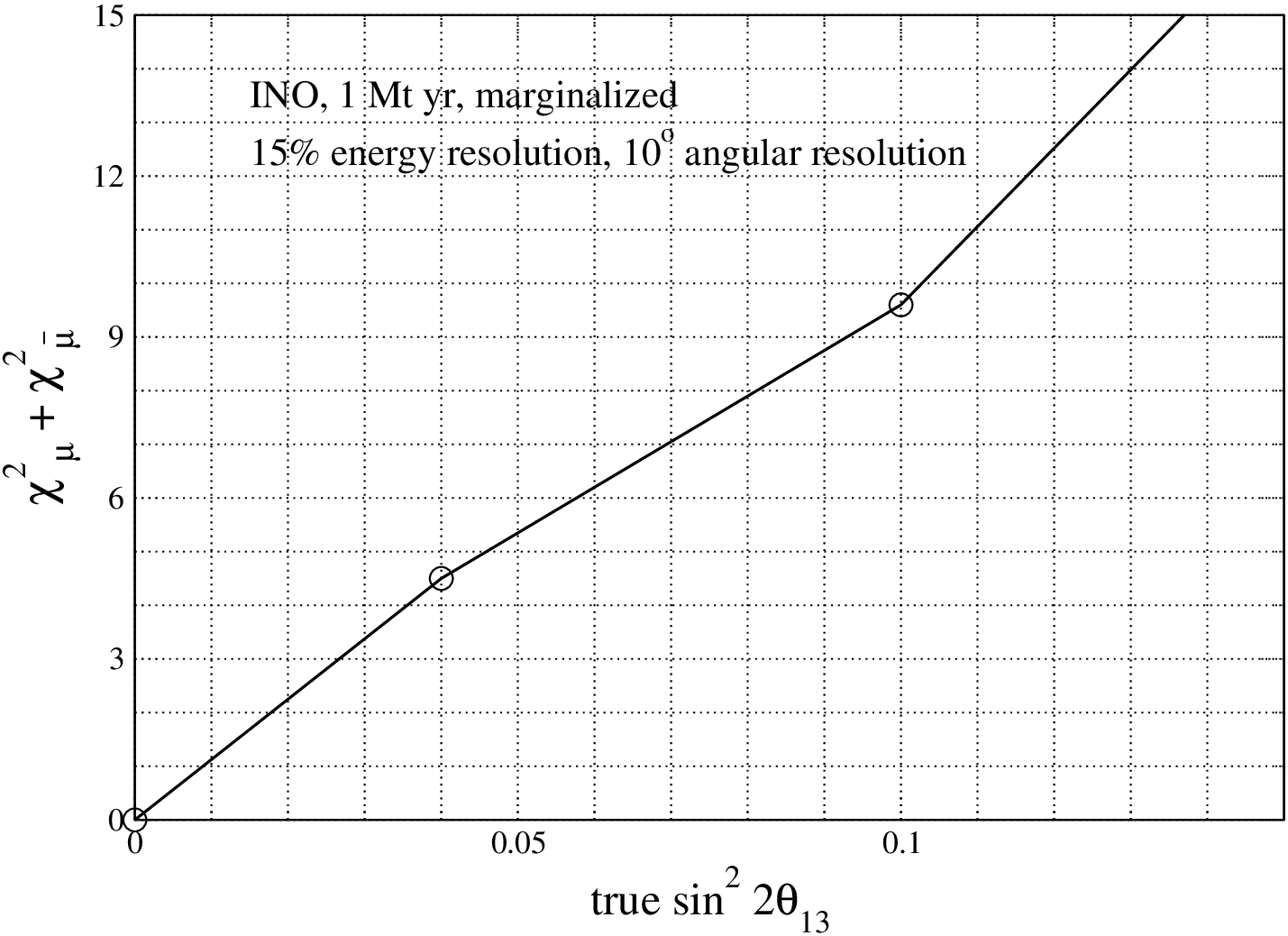}}
    \mycaption{Values of marginalized $\chi^2$ versus the input (true) value of $\sin^2 2 \theta_{13}$ for \INO (1
\mtyrd). Shown is the sum of $\chi^2_{\mu}$ and
$\chi^2_{\bar{\mu}}$.
An energy resolution of $15\%$ and an angular resolution of
$10^\circ$ are assumed.
}
    \label{fig14}
\end{figure*}


\section{Results}

The ability of an atmospheric neutrino detector to resolve the
hierarchy  depends on various factors.
We study the effect of these different factors by computing
the $\chi^2$ as a function of
\\
(a) energy and
angular resolution of the detector and
\\
(b) $\sin^2 2 \theta_{13}$, which controls the matter effects.
\\
In studying (a) the same
values of the neutrino parameters, $|\da|$, $\sin^2 2\theta_{23}$ and
$\sin^2 2 \theta_{13}$, are used in computing both $\rm{N^{ex}_{ij}}$
and $\rm{N^{th}_{ij}}$.
In studying (b)
we   marginalize over
these parameters.

For \HKd, the overall sensitivity is taken to be the sum of
$\chi^2$ values for the muon and electron event rates and  hence,
the total ${\mathrm{\chi^2 = \chi^2_{\mu+\bar{\mu}} +
\chi^2_{e+\bar{e}}}}$.

For the \INO analysis, the $\mu^-$ and $\mu^+$ event rate
are separately used to compute the $\chi^2$ sensitivity
to the hierarchy. The overall sensitivity is taken to be
the sum of $\chi^2$ values for the $\mu^-$ and $\mu^+$ event
rates. Hence, for \INOd, the total $\chi^2 = \chi^2_{\mu} +
\chi^2_{\bar{\mu}}$.

\subsection{Megaton water \chr detectors}

We first  discuss our results for megaton water \chr detectors.

The left and right panels of \Fig~{\ref{fig5}} give the values of
${\rm{\chi^2_{pull}}}$  from \eq~\ref{chisqpull} for muon events
($\chi^2_{\mu+{\bar{\mu}}}$) and for electron events
(${\mathrm{\chi^2_{e+{\bar{e}}}}}$) as a function of angular and
energy resolution respectively. In this figure all neutrino
parameters are kept fixed in both ${\rm N^{th}}$ and ${\rm
N^{ex}}$. Since the number of muon events is expected to be more
than that of the electron events, naively one would expect better
hierarchy discrimination in the muon channel. We find that this
indeed is the case when the energy and angular resolutions are
very good. But with worsening energy and angular resolutions, the
hierarchy resolving power of electron events becomes comparable to
that of muon events. This is primarily because the contribution to
the muon events comes from the probability $\pmumu$ and $\pemu$
whereas for electron events it is from $\pmue$ and $\pee$. As is
seen from the \Figs~\ref{fig1} and \ref{fig2}, the matter effects
in $\pmumu$ are generally smaller than those in $\pmue$ and
$\pee$. Also, the matter induced change in $\pmue$ and $\pee$ is
consistently of the same sign over all the relevant ranges in
energy and pathlength, while the matter induced change in $\pmumu$
is oscillatory for longer baselines. Worsening resolutions thus
have a stronger effect on muon events, and, over-riding their
statistical advantage, lead to ${\mathrm{\chi^2_{\mu+{\bar{\mu}}}
\leq \chi^2_{e+{\bar{e}}}}}$.

\Fig~\ref{fig6} and \Fig~\ref{fig7} give the values of the fixed
parameter $\chi^2_{\mu+{\bar{\mu}}}$ and
${\mathrm{\chi^2_{e+{\bar{e}}}}}$ as functions of the true value
of $\sin^2 2\theta_{13}$. In \Fig~\ref{fig6}, {\bf NH} is assumed
to be the true hierarchy, whereas in \Fig~\ref{fig7}, {\bf IH} is
assumed to be the true hierarchy. The $\chi^2$ values in the two
cases are quite close, as one would expect. Since $\theta_{13}$
drives the matter effects, it is no surprise that the $\chi^2$ is
larger for larger values of $\theta_{13}$. It is interesting to
examine the limit where  $\theta_{13}$ goes to zero for electron
and muon events. For the former, one needs to recall that the
relevant analytical expressions for $\pee$ and $\pemu$ in matter
are those which are exact in $\ds/\da$ and accurate to first order
in $\sin^2 2\theta_{13}$
~\cite{Freund:2001pn,Akhmedov:2004ny,Kimura:2002hb,Takamura:2005df}.
The leading order term in this expansion is devoid of any
hierarchy sensitivity, while the term to first order in $\sin^2
2\theta_{13}$ is zero. Thus we expect $\chi^2$ to go to zero in
this limit for electron events, and this is borne out by  the
right panels of both \Fig~\ref{fig6} and \Fig~\ref{fig7}. However,
we see from the left panels of these figures that at
$\theta_{13}=0$ the muon events have non-zero $\chi^2$. The muon
disappearance probability in case of $\theta_{13}=0$ can be
expressed as,
 \begin{eqnarray}
{\mathrm{ \pmumum = \left[~1 - \sin^2 2\theta_{23} ~\sin^2 \left\{
1.27~(1 -\alpha \;c^2_{12})  ~\da ~\frac{L}{E} \right\}~\right] }}
\end{eqnarray}
Here, $c_{12}$ denotes $\cos\theta_{12}$ and $\alpha$ is a
dimensionless parameter given by $\alpha=\ds/|\da|$. This gives us
 \be
{ \mathrm{ P_{\mu \mu}^{NH} - P_{\mu \mu}^{IH} = \sin^2
2\theta_{23} \;\left[\; \sin^2\left\{1.27(1+\alpha\;
c_{12}^2)\da\frac{L}{E}\right\} - \sin^2\left\{1.27(1 - \alpha
\;c_{12}^2)\da\frac{L}{E}\right\}\;\right] }}
\label{nh-ih-exact_strong} \ee
 Thus there is a hierarchy
sensitivity due to the term $\alpha=\ds/|\da|$ even for
$\theta_{13}=0$. This is true only if $|\da|$ is known very
precisely, permitting the use of the same fixed value of $|\da|$
for normal and inverted hierarchy. For current uncertainties in
$|\da|$, marginalization over this parameter leads to a wash out
of this sensitivity (as will be shown later). This happens when
the values of $|\da|$ in the computation of NH and IH
probabilities differ by $2 \ds
c_{12}^2$~\cite{deGouvea:2005hk,deGouvea:2005mi}. Hence we would
have a non-zero hierarchy sensitivity for $\theta_{13}=0$ only if
the error in the determination of $|\da|$ is less than the
magnitude of $\ds$.

Another interesting feature visible in \Fig~\ref{fig6} and
\Fig~\ref{fig7} is that for $\sin^2 2 \theta_{13} \geq 0.1$, the
${\mathrm{\chi^2_{e+{\bar{e}}}}}$ flattens out. As emphasized in
\Sec~3, the maximum difference between events in the case of NH
and those in the case of IH occurs for energies in the resonance
region. At resonance, the matter dependent mixing angle
${\mathrm{\theta^m_{13} \simeq \pi/2}}$, but the matter dependent
mass-squared difference $\dam$ takes its minimum value of $\da
\sin 2 \theta_{13}$. For intermediate  values of $\theta_{13}$,
(\ie~those that are not tiny but still significantly less than
0.1)
 this mass-squared difference is too small for
the oscillating term in $\pmuem$ and $\peem$ to come close to unity,
even for the largest pathlengths. But for $\sin^2 2 \theta_{13}
\geq 0.1$, the value of $\da \sin 2 \theta_{13}$ is
large enough such that (a) for a large range of pathlengths,
$\sin^2 (1.27 \da \sin2\theta_{13} L/E) \lsim 1$,
(b) $\pmuem$ is close to its maximum value, and
(c) $\peem$ is close to its minimum value. Thus, the difference
between electron events for NH and those for IH becomes
essentially independent of $\theta_{13}$ above a certain limit.

\Fig~{\ref{fig8}} gives the values of fixed parameter
$\chi^2_{\mu+{\bar{\mu}}}$ and ${\mathrm{\chi^2_{e+{\bar{e}}}}}$
as a function of the value of the CP phase $\dcp$ for muon and
electron events. This figure shows that the dependence of $\chi^2$
on $\dcp$ is mild. This occurs because the terms containing $\dcp$
in oscillation probabilities are also proportional to $\ds$, whose
effect on these probabilities is small. Therefore the exact value
of $\dcp$ has negligible effect on the hierarchy distinguishing
ability of megaton water \chr detectors.

Finally, we present results with marginalization over the neutrino
oscillation parameters $|\da|$, $\sin^2 2 \theta_{23}$ and $\sin^2
2\theta_{13}$. We add the $\chi^2$ of the muon events
($\chi^2_{\mu+{\bar{\mu}}}$) and that of the electron events
(${\mathrm{\chi^2_{e+{\bar{e}}}}}$) and compute the set of
$\chi^2$ where all the inputs in $\mathrm{N^{th}_{ij}}$ have been
varied over their allowed ranges. The minimum of this set, called
${\mathrm{\chi^2_{min}}}$ (\eq~\ref{chisqfinal}), is the quantity
characterizing the capability of \HK to distinguish the two
hierarchies. As explained earlier, this capability is a function
of the {\it true} value of $\theta_{13}$. In \Fig~\ref{fig9} we
plot the values of the marginalized
${\mathrm{\chi^2_{\mu+{\bar{\mu}}} + \chi^2_{e+{\bar{e}}}}}$ vs
the true value of $\sin^2 2\theta_{13}$. The $\chi^2$ curve is
flattened for $\sin^2 2 \theta_{13} \geq 0.1$ because of the
contribution of ${\mathrm{\chi^2_{e+{\bar{e}}}}}$, which, as
discussed earlier, flattens out for higher values of
$\theta_{13}$. We find the $\chi^2$ to be $\geq 4$ for $\sin^2 2
\theta_{13} \geq 0.05$. Thus if \DCHOOZ finds a non-zero value for
$\sin^2 2 \theta_{13}$, then the combination of muon and electron
events at megaton sized water \chr detectors can make a
distinction between the two hierarchies at $95\%$ CL over a $\sim
3$ year exposure period.
Note that, for $\sin^2 2 \theta_{13} = 0.05$,
the marginalized $\chi^2$ is about half of the fixed parameter $\chi^2$.

We have not included the backgrounds in our calculations. They
will be the same in both ``experimental" and ``theoretical"
spectra and cancel out in the numerator of ${\rm{\chi^2_{stat}}}$,
as can be seen from \eq~\ref{chisqstat}. They need to be included
in the denominator, which is the sum of ``signal" and background
events. Thus the denominator will be slightly larger than the
computed number $\rm{N^{ex}_{ij}}$ and ${\rm{\chi^2_{stat}}}$ will
be correspondingly lower. Thus the backgrounds can be adequately
taken care of if we divide ${\rm{\chi^2_{stat}}}$ of muon events
by {\bf 1.003} and that of electron events by {\bf 1.17}. We {\it
have not included} these factors in our calculations but their
inclusion makes a very small change in overall $\chi^2$.

\subsection
{Magnetized Iron detectors}

As stated previously, the iron detector which we use as
a prototype (based on the \INO design) has a mass of 100 kT
and is capable of separating the  $\nu_\mu$ and
$\bar{\nu}_\mu$ events, but is insensitive to any electron events.
We assume an exposure time
of 10 years and
first compute the fixed parameter ${\rm{\chi^2_{pull}}}$ from \eq~\ref{chisqpull}
and later the marginalized ${\rm{\chi^2_{min}}}$ (\eq~\ref{chisqfinal}).
In both cases, we compute $\chi^2_{\mu}$ and $\chi^2_{\bar{\mu}}$ separately and add them.
The  input values of neutrino parameters
are the same ones used in the case of water \chr detectors.

In \Fig~\ref{fig10}, we plot the fixed parameter $\chi^2_{\mu} + \chi^2_{\bar{\mu}}$ as a function of
angular resolution and energy resolution.
From the left panel, we observe that the $\chi^2$
increases sharply (by a factor of 3 or more) if the angular
resolution is improved from $10^\circ$ to $5^\circ$. This occurs
because the maxima and the minima of
$\pmumu$ are narrower in width and exhibit a more rapid
variation (compared to $\pee$) as the baseline changes,
as can be observed in \Figs~\ref{fig1} and
\ref{fig2}. The improvement in $\chi^2$ due to the improvement in
the energy resolution is more modest, as the right panel of
\Fig~\ref{fig10} shows. Thus it is imperative for such detectors
to improve their angular resolution to the best of their ability
~\cite{Petcov:2005rv}.

In \Fig~\ref{fig11}, we plot the fixed parameter $\chi^2_{\mu} + \chi^2_{\bar{\mu}}$
as a function of the input ({\it true}) value of $\sin^2 2 \theta_{13}$.
The left panel assumes NH  as the true hierarchy while
the right panel assumes  IH  as the true hierarchy. As
in the case of muon events in water \chr detectors, $\chi^2$
increases for increasing values of $\sin^2 2 \theta_{13}$. And
there is a small non-zero $\chi^2$ for $\theta_{13}=0$ because
of the effect of the non-zero value of
$\ds$ discussed earlier.

\Fig~\ref{fig13} demonstrates  the
variation of
$\chi^2_{\mu} + \chi^2_{\bar{\mu}}$ with $\dcp$ keeping the other parameters
fixed in both NH and IH.
As in the case of water \chr detectors, the $\chi^2$
in this case is also  insensitive of $\dcp$.

Finally, we marginalize over the neutrino oscillation parameters
$|\da|$, $\sin^2 2 \theta_{13}$ and $\sin^2 2 \theta_{23}$ and
present the marginalized $\chi^2_{\mu} + \chi^2_{\bar{\mu}}$ for
the magnetized iron detector in \Fig~\ref{fig14} as a function of
the input value of $\sin^2 2 \theta_{13}$. Here again, we find
that resolution of mass hierarchy at $\geq 95\%$ C.L. is possible
for values of $\sin^2 2 \theta_{13} \geq 0.05$. The $\chi^2$
demonstrates a steep rise with an increase in $\theta_{13}$ which
is due to the charge identification capability of magnetized iron
calorimeter detectors. For this type of detectors we add the
$\chi^2$ from muon events and that from anti-muon events, each of
which is a sensitive function of $\theta_{13}$, and therefore the
total $\chi^2$ has a large $\theta_{13}$ dependence. Note that the
$\chi^2$ for $\theta_{13}=0$ is no longer non-zero due to the
marginalization over $|\da|$.


%
\section{Summary}
%

\noindent In this paper, we have studied the hierarchy resolving
power of (a) megaton sized water \chr detectors (prototype: \HKd)
 and (b) magnetized iron detectors (prototype: \INOd).

The first class of detectors has two important advantages:
(a) a very large size leading to high statistics and
(b) the ability to detect  both muon and
electron events.
However, these detectors are insensitive to  the charge of the
lepton and therefore one needs to sum over the lepton and
antilepton events. The matter effect affects the lepton events for
NH and antilepton events for IH, and in the summed event rate
these effects can cancel each other. In actual practice however
the cancellation is only partial since the neutrino cross section
for producing leptons is about 2.5$-$3 times higher than the
corresponding anti-neutrino cross section. This leads to hierarchy
sensitivity in these detectors, and despite the above
disadvantages, the sheer weight of statistics can lead to
significant differences in the signal for the two hierarchies.

\begin{table}
\begin{center}
\begin{tabular}{c | c | c } \hline \hline &&
\\
\hspace{0.5cm}{\sf {$\sin^2 2 \theta_{13}$}}\hspace{0.5cm}
        & \hspace{0.5cm}{\sf {$\chi^2_{\mu + {\bar{\mu}}} + \chi^2_{e + {\bar{e}}}$ (\HKd)}}\hspace{0.5cm} & \hspace{0.5cm}{\sf { $\chi^2_{\mu} + \chi^2_{\bar{\mu}}$ (\INOd)}}\hspace{0.5cm}
        \\ && \\
        \hline \hline
        0.0 &  0.0  &  0.0  \\
        0.04  &  3.6 &  4.5  \\
        0.10  &  5.9 &  9.6  \\
        0.15  &  7.1 &  16.9
          \\
          \hline \hline
\end{tabular}
\mycaption{Values of total marginalized $\chi^2$ with pull and priors, for \HK (1.8 \mtyrd) and \INO (1 \mtyrd).}
\label{HKINO}
\end{center}
\end{table}

For the binned $\chi^2$ analysis of atmospheric neutrino data it
is important to include the effect of energy and angular smearing.
This effect is more significant for muon events than for electron
events. This is because the matter effects in $\pmumu$ are
generally smaller than those in $\pee$ and $\pmue$ over the energy
and baseline ranges under consideration here. Also, $\pmumu$
exhibits an oscillatory behaviour with energy at the longer
baseline values ($\sim$ 10000 km) which tends to get washed out
with smearing. $\pee$ and $\pmue$, on the other hand, have a
significant matter effect over a broad range of energies and
pathlengths, which is less affected by smearing. Thus the electron
events gain in significance, and the cumulative effect from a
number of bins adds up to create a large difference between the
hierarchies, even though the electron neutrino flux is only about
half of the muon neutrino flux.
Together, the differences induced in $\mu$-like and e-like events
lead to a 2$\sigma$ signal for neutrino mass hierarchy for
$\sin^22\theta_{13} \approx 0.04$ for a moderate exposure time of
3.3 years. For larger exposure times, it is possible to determine
hierarchy at a higher statistical significance.

Magnetized iron calorimeter detectors, on the other hand, are
sensitive only to muons. But the magnetic field endows them with
charge identification capability. Therefore one can collect the
muon and the anti-muon events separately and compute the $\chi^2$
for each type of events and then add these. Thus the hierarchy
sensitivity is considerably enhanced as compared to water \chr
detectors, and similar statistical significance can be achieved
with smaller statistics.

In \Tab~\ref{HKINO} we compare the $\chi^2$ sensitivity of both
type of detectors for different values of $\sin^2 2\theta_{13}$.
We find that an exposure of about 2 \mtyr for water \chr and 1
\mtyr for magnetized iron detectors can resolve the matter
hierarchy at $95\%$ C.L. or better, provided $\sin^2 2 \theta_{13}
\geq 0.05$. This table and \Figs~\ref{fig9} and \ref{fig14}
summarize our main results, where ${\mathrm{\chi^2_{min}}}$ is
plotted versus the input value of $\stsmall$. This
${\mathrm{\chi^2_{min}}}$ incorporates theoretical, statistical
and systematic errors, along with smearing over neutrino energy
and direction and also marginalization over the allowed ranges of
neutrino parameters. These figures as well as the table show that
the variation of $\chi^2$ with increasing $\sin^2 2\theta_{13}$
for water \chr detectors is significantly flatter as compared to
that for magnetized iron calorimeter detectors. This is due to (a)
The contribution of ${\mathrm{\chi^2_{e+{\bar{e}}}}}$ in the
$\chi^2$ for \HKd, which flattens out for higher values of
$\theta_{13}$; and (b) $\chi^2_{\mu+{\bar{\mu}}}$ is a relatively
less sensitive function of $\theta_{13}$ since the addition of
muon and anti-muon events partially cancels the $\theta_{13}$
dependence in the difference in NH and IH events
(\eq~\ref{eq:Deltan}). On the other hand, for magnetized iron
calorimeter detectors $\chi^2$ is the sum of the $\chi^2$ from
muon events and that from anti-muon events and the $\theta_{13}$
sensitivity adds constructively in this case.

The $\chi^2$ for muon events exhibits a higher sensitivity to
improved angular and energy resolution than do electron events.
Hence, there is a dramatic improvement in the hierarchy
discrimination capability of magnetized iron detectors with
improved angular resolution. The results quoted in the previous
paragraph assume a modest energy resolution of $15\%$ and angular
resolution of $10^\circ$. Increasing the the angular resolution
from $10^\circ$ to $5^\circ$ dramatically increases the $\chi^2$
by a factor of $3$. Improvement in the energy resolution for a
magnetized detector leads to a more modest improvement of the
$\chi^2$. Thus it is imperative for these  detectors to have good
angular resolution if they are to resolve the mass hierarchy at a
statistically significant level.

\subsection*{Acknowledgements} We would like to thank S. Choubey for
helpful discussions and cross-checking our numerical code. We
thank J. Kopp, T. Schwetz and T. Kajita for useful discussions.
S.G. was supported by the Alexander von-Humboldt Foundation during
a part of this work and would like to thank Max-Planck-Institut
for Kernphysik, Heidelberg and Tata Institute of Fundamental
Research, Mumbai for hospitality during the final phases of this
work. R.G. would like to acknowledge the support and hospitality
of the Aspen Center for Physics and the SLAC theory group while
the final stages of this work were in progress. S.U.S. would like
to thank the theory group at CERN for their hospitality during the
finishing stages of this work. S.G. and S.U.S. thank BRNS project
number 2006/37/9 (Govt. of India) for partial financial support.
S.S. would like to thank the \INO collaboration for support. P.M.
acknowledges the Weizmann Institute of Science, Israel for
financial support. Part of the computational work for this study
was carried out at cluster computing facility in the
Harish-Chandra Research Institute
(http://www.cluster.mri.ernet.in).


\begin{appendix}

\renewcommand{\theequation}{\thesection\arabic{equation}}

\setcounter{equation}{0}
\section{Calculation of the normalization constant}
\label{app}


In this appendix, we give the procedure for computing the Gaussian
resolution function for angular smearing.

The flux distribution in terms of the zenith direction is given as
${\rm{(d \Phi/d \cos \theta_t)}}$.
Thus the flux per solid angle is given by ${\rm{(1/2 \pi) (d \Phi/d \cos
\theta_t)}}$. The number of events per unit solid angle are given by
\begin{equation}
{\mathrm{
\frac{d N_{\mu}}{d \Omega_t} =
{\rm{\frac{1}{2 \pi} ~ \left[
\left(\frac{d \Phi}{d \cos \theta_t} \right)_{\mu} ~P_{\mu \mu} +
\left(\frac{d \Phi}{d \cos \theta_t} \right)_{e} ~P_{e \mu}
\right]}}~\sigma_{CC}
}}
\label{truedist}
\end{equation}
Even though the left hand side says ${\mathrm{\Omega_t}}$, this
quantity is independent of ${\mathrm{\phi_t}}$. This plays a key
role in later calculation. In the above expression, the
disappearance and appearance
probabilities are functions of neutrino travel distance and hence of ${\mathrm{\theta_t}}$.
In addition, they and the cross section are also functions of
neutrino energy, which we will not explicitly consider in this discussion.

Now consider a unit sphere with two nearby points whose coordinates are
${\mathrm{(\theta_t,\phi_t)}}$ and ${\mathrm{(\theta_m,\phi_m)}}$.
The distance between them is given by
\begin{equation}
{\mathrm{ds^2
= (\theta_t - \theta_m)^2 + \sin^2 \theta_t ~(\phi_t -\phi_m)^2}}
\end{equation}
Let $\Delta \theta$ be the uncertainty in the determination of the
neutrino direction, which is given by the point ${\mathrm{(\theta_t,\phi_t)}}$.
Then the smearing factor for the solid angle is
\begin{equation}
{\mathrm{R(\Omega_t,\Omega_m) = N  \exp \left[ -~ \frac{(\theta_t
- \theta_m)^2 + \sin^2 \theta_t ~(\phi_t - \phi_m)^2}{2 ~(\Delta
\theta)^2} \right]}} \label{smearfacnew}
\end{equation}
Here, the normalization factor ${\mathrm{N = 1/{\rm{A(\theta_t)}}}}$, where ${\mathrm{A(\theta_t)}}$ is the area of the Gaussian for each value of ${\rm{\theta_t}}$. Since the Gaussians at the edges are truncated, this causes a fall in the area of the side bins if the normalization factor used is $\sqrt{2\pi}\sigma$ (the area of a full Gaussian). We require the total event number to be preserved when the range of ${\mathrm{\cos \theta_m}}$ is the full range (-1 to +1), since any value of $\cos \theta$ outside this range is unphysical and hence no events should come in or be thrown out. So, the normalization has to be by the actual area of each Gaussian.

For each value of ${\mathrm{\theta_t}}$, the Gaussian area is given by
\begin{equation}
{\mathrm{A(\theta_t)}}  = {\mathrm{\int_{-\pi}^{\pi} d \phi_m ~\int_{0}^{\pi} d \theta_m ~\sin \theta_m
~\exp \left[ -~ \frac{(\theta_t - \theta_m)^2 +
\sin^2 \theta_t ~(\phi_t - \phi_m)^2}{2~ (\Delta \theta)^2} \right]}}
\label{area}
\end{equation}
On computing the ${\mathrm{\phi_m}}$ integral, using the same method
described below in \eq~\ref{phim1}-\ref{phim3}, we get
\begin{equation}
{\mathrm{N(\theta_t) = \frac{1}{A(\theta_t)} = \left[\int_{-1}^{1} d \cos \theta_m~
\exp \left[ -~ \frac{(\theta_t - \theta_m)^2}{2 (\Delta \theta)^2} \right]
\frac{\sqrt{2\pi} ~\Delta \theta}{\sin \theta_t}\right]^{-1}
}}
\label{norm}
\end{equation}
Combining the smearing factor from \eq~\ref{smearfac} with the
distribution in true angles coming from \eq~\ref{truedist}, and
integrating over ${\mathrm{\phi_t}}$ and ${\mathrm{\theta_t}}$ we get the distribution
in measured angles
\begin{equation}
{\mathrm{\frac{d N_\mu}{d \Omega_m} = \int \left(\frac{d N_\mu }{d \Omega_t}\right) ~R(\Omega_t, \Omega_m)
~d \Omega_t}}
\end{equation}
where ${\mathrm{d \Omega_t = d \theta_t \sin \theta_t d \phi_t}}$. In the above
equation, ${\mathrm{\phi_t}}$ appears only in the smearing factor and it appears
only as ${\mathrm{(\phi_t - \phi_m)^2}}$. We can define this as ${\mathrm{\tilde{\phi}^2_m}}$
so that the integrand is completely independent of ${\mathrm{\phi_t}}$ and that
integration is trivial. Thus the distribution in measured angles is
given by
\begin{equation}
{\mathrm{\frac{d N_\mu }{d \Omega_m} =
(2 \pi)~ \int_{0}^{\pi}~ \left(\frac{d N_\mu }{d \Omega_t}\right)
~R(\Omega_t, \Omega_m) ~\sin \theta_t ~d \theta_t
}}
\end{equation}

Now we do the integral over ${\mathrm{\phi_m}}$ so that we have the distribution
over ${\mathrm{\cos \theta_m}}$ on LHS and ${\rm{\cos \theta_t}}$ on RHS. On RHS, only
the smearing factor ${\mathrm{R(\Omega_t,\Omega_m)}}$ depends on ${\mathrm{\phi_m}}$.
Thus doing the ${\mathrm{\phi_m}}$ integral we get
\begin{eqnarray}
{\mathrm{\int \frac{d N_\mu}{d \Omega_m} ~d \phi_m  =
2 \pi
\int_{-\pi}^{\pi}
\int_{0}^{\pi}
~\frac{d N_\mu}{d \Omega_t}
~N(\theta_t)
~\exp \left[ -~ \frac{(\theta_t - \theta_m)^2 +
\sin^2 \theta_t ~(\phi_t - \phi_m)^2}
{2 ~(\Delta \theta)^2}
\right]
\sin \theta_t ~d \theta_t ~d \phi_m}}
\nonumber\\
\label{phim1}
\end{eqnarray}
Changing variable to $\tilde{\phi}_m = \phi_m - \phi_t$, we get
\begin{eqnarray}
{\mathrm{\int \frac{d N_\mu}{d \Omega_m} ~d \phi_m =
2 \pi ~\int_{-\pi}^{\pi} ~\int_{0}^{\pi} ~\frac{d N_\mu}{d \Omega_t}
 ~N(\theta_t)
~\exp \left[ -~ \frac{(\theta_t - \theta_m)^2 +
\sin^2 \theta_t ~(\tilde{\phi}_m)^2}{2 ~(\Delta \theta)^2} \right]
\sin \theta_t ~d \theta_t ~d \tilde{\phi}_m}}
\nonumber\\
\label{phim2}
\end{eqnarray}
In doing the ${\rm{\phi_m}}$ (${\rm{\tilde{\phi}_m}}$) integration, we extend the
limits of integration from $\pm \pi$ to $\pm \infty$. This leads to
negligible error because the Gaussian falls off very steeply. After
the integration, we get
\begin{eqnarray}
{\mathrm{\frac{d N_\mu}{d \cos \theta_m} =
2 \pi \int_{0}^{\pi} \frac{d N_\mu}{d \Omega_t}
~ N(\theta_t)
~\exp \left[ -~\frac{(\theta_t - \theta_m)^2}{2 (\Delta \theta)^2} \right]
~\sin \theta_t ~d \theta_t ~\sqrt{2 \pi} ~\Delta \theta ~\frac{1}{\sin \theta_t}}}
\label{phim3}
\end{eqnarray}
We can cancel one factor of ${\rm{\sin \theta_t}}$
on the RHS. On substituting for ${\rm{N(\theta_t)}}$ from \eq~\ref{norm}
and ${\rm{d N_\mu/ d \Omega_t}}$ from \eq~\ref{truedist},
we get the binned distribution in ${\rm{\cos \theta_m}}$ to be
\begin{eqnarray}
{\rm{d N_\mu}} = {\rm{ \int d\theta_t \frac{
 [(d \Phi/d \cos \theta_t)_{\mu} P_{\mu\mu} + (d \Phi/d \cos \theta_t)_{e} P_{e\mu}] \sigma_{CC}
\int_{bin} d\cos \theta_m
\exp \left[ -~ \frac{(\theta_t - \theta_m)^2}{2 ~(\Delta \theta)^2} \right]
\sqrt{2 \pi} \Delta \theta }
{\int_{-1}^{1} d\cos \theta_m
\exp \left[ -~ \frac{(\theta_t - \theta_m)^2}{2 ~(\Delta \theta)^2} \right]
\frac{\sqrt{2\pi} \Delta \theta}{\sin\ \theta_t}}}}
\nonumber \\
= {\rm{\int d\theta_t \frac{
 [(d \Phi/d \cos \theta_t)_{\mu} P_{\mu\mu} + (d \Phi/d \cos \theta_t)_{e} P_{e\mu}] \sigma_{CC}
\sin \theta_t
\int_{bin} d\cos \theta_m
\exp \left[ - \frac{(\theta_t - \theta_m)^2}{2 ~(\Delta \theta)^2} \right]
}{\int_{-1}^{1} d\cos \theta_m
\exp \left[ - \frac{(\theta_t - \theta_m)^2}{2 ~(\Delta \theta)^2} \right]}}
}
\nonumber\\
\end{eqnarray}

\end{appendix}

\bibliographystyle{apsrevwinter}

\bibliography{myrefsept}

\end{document}